\documentclass[usenatbib]{mnras}

\usepackage{newtxtext,newtxmath}

\usepackage[T1]{fontenc}
\usepackage{amsmath}

\DeclareRobustCommand{\VAN}[3]{#2}
\let\VANthebibliography\thebibliography
\def\thebibliography{\DeclareRobustCommand{\VAN}[3]{##3}\VANthebibliography}

\usepackage{ae,aecompl}
\usepackage{xparse,xspace}
\usepackage{cancel}
\usepackage{siunitx}
\usepackage{graphicx}	
\usepackage{amsmath}	
\usepackage{amsfonts}
\usepackage{graphicx}
\usepackage{color}
\usepackage{bm}
\usepackage{lipsum}
\usepackage{nccmath}
\usepackage{tabularx}
\usepackage{caption}
\usepackage{multicol}
\usepackage{graphicx}
\usepackage[export]{adjustbox}
\usepackage[flushleft]{threeparttable}

\title[The early evolution of magnetar rotation - II]{The early evolution of magnetar rotation - II. Rapidly rotating magnetars: Implications for Gamma-Ray Bursts and Super Luminous Supernovae}

\author[Prasanna et al.]{
Tejas Prasanna$^{1,2}$\thanks{E-mail: prasanna.9@osu.edu},
Matthew S. B. Coleman$^{3,4}$\thanks{E-mail: matthew.s.b.coleman@gmail.com},
Matthias J. Raives$^{2,5,6}$\thanks{E-mail: mraives@carnegiescience.edu},
\& Todd A. Thompson$^{2,5,1}$\thanks{E-mail: thompson.1847@osu.edu}
\\
$^{1}$Department of Physics, The Ohio State University, Columbus, Ohio 43210, USA\\
$^{2}$Center for Cosmology \& Astro-Particle Physics, The Ohio State University, Columbus, Ohio 43210, USA\\
$^{3}$Department of Astrophysical Sciences, Princeton, NJ 08540 USA\\
$^{4}$Department of Physics and Engineering Physics, Stevens Institute of Technology,
Castle Point on the Hudson, Hoboken, NJ 07030, USA\\
$^{5}$Department of Astronomy, The Ohio State University, Columbus, Ohio 43210, USA\\
$^{6}$The Observatories of the Carnegie Institution for Science, 813 Santa Barbara St., Pasadena, CA 91101, USA\\
}

\date{Accepted XXX. Received YYY; in original form ZZZ}

\pubyear{2023}

\begin{document}

\label{firstpage}
\pagerange{\pageref{firstpage}--\pageref{lastpage}}
\maketitle

\begin{abstract}
Rapidly rotating magnetars have been associated with gamma-ray bursts (GRBs) and super-luminous supernovae (SLSNe). Using a suite of 2D magnetohydrodynamic simulations at fixed neutrino luminosity and a couple of evolutionary models with evolving neutrino luminosity and magnetar spin period, we show that magnetars are viable central engines for powering GRBs and SLSNe. We also present analytic estimates of the energy outflow rate from the proto-neutron star (PNS) as a function of polar magnetic field strength $B_0$,  PNS angular velocity $\Omega_{\star}$, PNS radius $R_{\star}$ and mass outflow rate $\dot{M}$. We show that rapidly rotating magnetars with spin periods $P_{\star}\lesssim 4$\,ms and polar magnetic field strength $B_0\gtrsim 10^{15}$\,G can release $10^{50}-5\times 10^{51}$\,ergs of energy during the first $\sim2$\,s of the cooling phase. Based on this result, it is plausible that sustained energy injection by magnetars through the relativistic wind phase can power GRBs. We also show that magnetars with moderate field strengths of $B_0\lesssim 5\times 10^{14}$\,G do not release a large fraction of their rotational kinetic energy during the cooling phase and hence, are not likely to power GRBs. Although we cannot simulate to times greater than $\sim 3-5$\,s after a supernova, we can hypothesize that moderate field strength magnetars can brighten the supernova light curves by releasing their rotational kinetic energy via magnetic dipole radiation on timescales of days to weeks, since these do not expend most of their rotational kinetic energy during the early cooling phase.
\end{abstract}



\section{Introduction}
\label{section:introduction}
 The mechanism of gamma-ray bursts (GRBs) has been a critical area of research since their discovery \citep{Klebesadel1973}. Some long-duration GRBs are known to be accompanied by core collapse supernovae (SNe) (e.g. \citealt{Stanek2003, Woosley2006}), suggesting a central engine that could either be a stellar-mass black hole (a ``collapsar,'' e.g.,  \citealt{Woosley1993, MacFadyen1999, Barkov2008, Gottlieb2022}) or a rapidly-rotating, highly-magnetic neutron star  (a ``millisecond magnetar,'' e.g., \citealt{Usov1992, Thompson2004, Metzger2007, Metzger2011}). Magnetar models are also invoked for short-duration GRBs with extended emission \citep{Metzger2008, Bucciantini2012} and GRBs having X-ray plateaus (e.g. \citealt{Lyons2010, Rowlinson2014}). However, some studies suggest that short duration GRBs may not be produced by long-lived magnetars (see \citealt{Beniamini2021, Ricci2021} for example). \cite{Ricci2021} argue that a powerful magnetar as a merger remnant can be excluded for certain range of ejecta mass.
 
 In the millisecond magnetar model, a strongly magnetized proto-neutron star (PNS) is born in massive star collapse \citep{Duncan1992}, or in the merger of a white dwarf binary \citep{King2001, Levan2006, Metzger2008}, or via the accretion induced collapse of a white dwarf \citep{Nomoto1991, Metzger2008}. The strong magnetic field may originate from flux-freezing during the collapse of a strongly-magnetized core or via an efficient dynamo \citep{Duncan1992,Thompson1993,White2022}. In the first few seconds after formation, the PNS drives a non-relativistic neutrino-heated wind \citep{Duncan1986} that may be the site for production of some $r$- and $p$-process nuclei \citep{Woosley1993,Janka1996,Otsuki2000,Thompson2001,Wanajo2001,Wanajo2006, Pruet2006}. Even with the high magnetic field strengths of magnetars, the wind at these early times is non-relativistic, but can be magneto-centrifugally accelerated if the rotation rate is high enough \citep{Thompson2004}. As the neutrino luminosity decreases on few-second timescales, the mass loss rate decreases, the wind becomes increasingly magnetically-dominated \citep{Thompson2003}, the Alfv\'en surface approaches the light cylinder, and the flow transitions to a relativistic, Poynting flux-dominated regime. As the PNS neutrino cooling epoch ends and the magnetization of the flow becomes very large, continued magnetar spindown is powered by inefficient classical ``vacuum'' dipole radiation (e.g., \citealt{Usov1992,Zhang2001}). Throughout, energy and angular momentum are extracted from the rotating PNS. During the non-relativistic and semi-relativistic phases, the spindown power can be many orders of magnitude larger than what would be expected from dipole spindown alone \citep{Thompson2004,Prasanna2022}. \cite{Metzger2011} produce 1-zone evolutionary models of cooling millisecond proto-magnetars and argue that their winds can consistently account for the observed properties of GRBs, depending on the initial magnetic field strength and spin period, and on assumptions about the GRB emission mechanism. Further, \cite{Bucciantini2007}, \cite{Komissarov2007}, and \cite{Bucciantini2008} show that the energy extracted from the proto-magnetar can in principal be channeled into a highly-collimated jet that subsequently produces a GRB. When the magnetic and rotation axes are misaligned, there may be significant dissipation in the equatorial current sheet region \citep{Margalit2018}. The early magneto-centrifugal acceleration of PNS winds may also be critical to our understanding of GW170817 and other neutron star binary mergers \citep{Metzger2018,Mosta2020}. There are however a number of studies that argue against a long-lived magnetar central engine for GW170817 (see for example \citealt{Granot2017, Van_putten2023}).

In magnetar models of GRBs, the magnetic field strength needs to be $10^{15}-10^{16}$\,G so that the rotational energy is extracted on few-second timescales. Application of the magnetar model to super-luminous supernovae (SLSNe) implies central engines with moderate field strengths of $\sim 10^{14}$\,G, lower than those required for GRBs. The more modest magnetic field strengths required by magnetar models of SLSNe reflect the fact that for a large increase in SN luminosity, the spindown time and the photon diffusion time in the envelope should be comparable, and typically days to weeks \citep{Kasen2010,Woosley2010}. Magnetars born with field strengths $\sim10^{14}$\,G and with initial rotation periods of $\sim 2-20$\,ms can explain some of the brightest events observed with luminosity $\gtrsim 10^{44}$\,ergs\,s$^{-1}$ \citep{Kasen2010, Nicholl2017}.

Here, we present two-dimensional axisymmetric models of rapidly-rotating proto-magnetars during their early few-second non-relativistic magneto-centrifugal wind epoch in the range of parameters relevant to both GRBs and SLSNe. Early semi-analytic and numerical models of proto-magnetar winds in this phase were one-dimensional \citep{Thompson2004,Metzger2007}. Previous two-dimensional relativistic simulations by \cite{Bucciantini2007,Bucciantini2008} explored the wind dynamics and spindown rate using a parameterized equation of state with no neutrino heating/cooling. Still other works have explored the neutrino-heated dynamics and nucleosynthesis in static, assumed magnetic field geometries \citep{Vlasov2014,Vlasov2017}.  Other works like \cite{Dessart2008} and \cite{Mosta2014} focus on the collapse, explosion, and early evolution of jets from the neutron star, but they do not evolve to long timescales to study the wind dynamics during the cooling epoch and for a variety of magnetic field strengths and initial PNS spin periods. Other recent works explore the structure of rapidly rotating PNSs and their winds, but without magnetic fields \citep{Desai2022}.  An important early work analogous to ours here is by \cite{Komissarov2007}, who present 2d relativistic axisymmetric winds with neutrino heating/cooling and a general EOS, but do not explore a wide range of initial parameters.

In this paper, we compute wind models with polar magnetic field strength $B_0$ ranging from $10^{13}-3\times10^{15}$\,G and initial rotation periods $P_{\star}<10$\,ms representative of rotation periods and field strengths required to explain GRBs and SLSNe. We focus on the energy outflow from the PNS during the first $\sim2$\,s of the cooling phase. We also present results for initial magnetar spin periods of 16\,ms and 32\,ms to connect with our previous work on slowly rotating magnetars \citep{Prasanna2022}, where we showed that magnetars born with initial spin periods $\gtrsim100$\,ms, representative of the normal pulsar population, can spin down very rapidly during the PNS cooling epoch. Known magnetars have spin periods of a few to ten seconds with some extreme systems having longer spin periods \citep{Olausen2014, Luca2006}. As discussed in \cite{Prasanna2022}, some magnetars have a characteristic age (determined using the standard magnetic dipole formula) which is longer than the respective supernova remnant age, suggesting that strong early spindown maybe important for some magnetars. In Section \ref{section:model}, we discuss the numerical technique, microphysics employed and boundary conditions. In Section \ref{results}, we present the results from our 2D simulations with focus on energy outflow rate $\dot{E}$ from the PNS as a function $B_0$ and $P_{\star}$. In Section \ref{results}, we also provide analytic estimates of $\dot{E}$ closely following \cite{Thompson2004} and \cite{Metzger2007}. In Section \ref{conclusions}, we discuss the implications of energy injection by the PNS onto the surroundings and associate magnetars with GRBs and SLSNe.  

\section{Model}
\label{section:model}
We use the MHD code Athena++ \citep{Athena++} for our simulations, which we have configured to solve the non-relativistic magneto-hydrodynamic (MHD) equations:
\begin{align}
          \frac{\partial \rho}{\partial t} + \nabla\cdot\left(\rho\mathbfit{v}\right)&=0,\label{eq:continuity}\\
          \frac{\partial \left(\rho\mathbfit{v}\right)}{\partial t} + \nabla\cdot\left[\rho\mathbfit{vv}+\left(P+\dfrac{B^2}{2}\right)\mathbf{I}-\mathbfit{BB}\right]&=-\rho \frac{GM_{\star}}{r^2}\boldsymbol{\hat{r}},\label{eq:momentum}\\
          \frac{\partial E}{\partial t} + \nabla\cdot\left[\left(E+\left(P+\dfrac{B^2}{2}\right)\right)\mathbfit{v}-\mathbfit{B}\left(\mathbfit{B}\cdot\mathbfit{v}\right)\right]&=\dot{Q},\label{eq:energy}\\
          \frac{\partial \mathbfit{B}}{\partial t} -\nabla\times\left(\mathbfit{v}\times\mathbfit{B}\right)&=0,\label{eq:eulermag}
        \end{align}
where $M_{\star}$ is the mass of the PNS, $r$ is the radius from the center of the PNS, $\rho$ is the mass density of the fluid, $\mathbfit{v}$ is the fluid velocity, $E$ is the total energy density of the fluid, $P$ is the fluid pressure, $\dot{Q}$ is the neutrino heating/cooling rate (defined in the next Section, see also \citealt{Prasanna2022}), and $\mathbfit{B}$ is the magnetic field.

\subsection{Microphysics}
\label{micro}
We employ the general equation of state (EOS) module in Athena++ \citep{Coleman2020}\footnote{With additional modifications to allow for a composition dependent EOS.} and use the approximate analytic form of the general EOS \citep{QW1996} containing non-relativistic baryons, relativistic electrons and positrons, and photons. The neutrino heating and cooling rates include charged-current neutrino absorption and electron capture on free nucleons \citep{Prasanna2022}. Different from \cite{Prasanna2022}, in this work we explicitly evolve the electron fraction $Y_{\rm e}$ as a function of radius and time. The electron fraction is updated at every radius at each timestep through a source function in Athena++. The EOS is dependent on the values of $Y_{\rm e}$ and has access to the updated values of $Y_{\rm e}$ at each timestep. The electron fraction is evolved using the absorption rates of neutrinos, electrons and positrons on nucleons. The rates in units of s$^{-1}$ are \citep{QW1996}:
\begin{equation}
\label{for1}
    \lambda_{\nu_{\rm e} n}\approx \frac{1+3\alpha^2}{2\pi^2}\left(\hbar c\right)^2G_{\rm F}^2 \frac{L_{\rm \nu_e}}{R_{\star}^2}\left(\frac{\langle\epsilon_{\rm \nu_e}^2\rangle}{\langle\epsilon_{\rm \nu_e}\rangle} +2\Delta + \frac{\Delta^2}{\langle\epsilon_{\rm \nu_e}\rangle}\right)(1-x),
\end{equation}
\begin{equation}
\label{for2}
    \lambda_{\bar{\nu}_{\rm e} n}\approx \frac{1+3\alpha^2}{2\pi^2}\left(\hbar c\right)^2G_{\rm F}^2 \frac{L_{\rm \bar{\nu}_e}}{R_{\star}^2}\left(\frac{\langle\epsilon_{\rm \bar{\nu}_e}^2\rangle}{\langle\epsilon_{\rm \bar{\nu}_e}\rangle} -2\Delta + \frac{\Delta^2}{\langle\epsilon_{\rm \bar{\nu}_e}\rangle}\right)(1-x),
\end{equation}
\begin{equation}
\label{rev1}
    \lambda_{e^{-}p} \approx 0.448 \ T_{\rm MeV}^5 \  \frac{F_4(\eta_{\rm e})}{F_4(0)},
\end{equation}
\begin{equation}
\label{rev2}
    \lambda_{e^{+}n} \approx 0.448 \ T_{\rm MeV}^5 \  \frac{F_4(-\eta_{\rm e})}{F_4(0)},
\end{equation}
where $\alpha=1.254$, $\Delta=1.2935$\,MeV is the neutron-proton mass difference, $\hbar$ is the reduced Planck's constant, $c$ is the speed of light, $G_{\rm F}$ is the reduced Fermi coupling constant which denotes the strength of interaction, $\langle \epsilon_{\nu}^n \rangle$ is the $n$-th neutrino energy moment, $L_{\nu}$ is the neutrino luminosity, $T_{\rm MeV}$ is the fluid temperature in MeV, $\eta_{\rm e}=\mu_{\rm e}/kT$ is the electron degeneracy parameter with $\mu_{\rm e}$ being the electron chemical potential and $x=\sqrt{1-R_{\star}^2/r^2}$ with $R_{\star}$ being the radius of PNS. We define the neutrino energy moments analogous to \cite{Scheck2006}:
\begin{equation}
    \langle \epsilon_{\nu}^n \rangle = \left(k_{\rm B}T_{\nu}\right)^n \frac{F_{n+2}(0)}{F_{2}(0)},
\end{equation}
where $k_{\rm B}$ in the Boltzmann constant, $T_{\nu}$ is the spectral temperature and $F_n$ is the Fermi integral. We assume that the degeneracy parameter is zero for neutrinos. We assume that the mean energies $\langle\epsilon_{\rm \bar{\nu}_e}\rangle=14$\,MeV and $\langle\epsilon_{\rm \nu_e}\rangle=11$\,MeV remain constant during the first $\sim3$\,s of the PNS cooling phase (see Section \ref{results} and \citealt{Pons1999, Prasanna2022}). Note that \cite{QW1996} assume $\eta_{\rm e}=0$ for the rates in equations \ref{rev1} and \ref{rev2}, but we find that it is important to include the Fermi integrals in order to correctly set $Y_{\rm e}$ near the PNS surface. $Y_{\rm e}\ll0.1$ at the surface and reaches asymptotic value within $\sim 2$ proto-neutron star (PNS) radii. For spin periods $P_{\star}\leq 3$\,ms, it is important to account for small $Y_{\rm e}$ near the surface in order to correctly set the temperature at the surface and enforce net zero neutrino heating in the first active zone of the computational grid (boundary conditions are explained in the following Section). We note that a smaller $Y_{\rm e}$ decreases the neutrino heating term at the surface (see eqn. 10 in \citealt{QW1996}). Assuming a constant $Y_{\rm e}$ with radius as in \cite{Prasanna2022} slightly overestimates spindown and the energy outflow rate $\dot{E}$ (see Table \ref{table1} and Section \ref{results}). For example, starting from an initial spin period of 200\,ms, at a resolution (see Section \ref{res}) of $(N_r\,,\,N_{\theta})=(256\,,\,128)$ and at a polar magnetic field strength of $B_0=2\times10^{15}$\,G, the PNS spins down to a period of 418\,ms assuming a constant $Y_{\rm e}=0.45$, while the PNS spins down to a period of 342\,ms with $Y_{\rm e}$ evolving as a function of time and radius at the end of 3\,s of evolution (see \citealt{Prasanna2022} for the details of spin period evolution of slowly rotating magnetars). At a spin period of 4\,ms, $B_0=10^{15}$\,G and electron anti-neutrino luminosity $L_{\rm \bar{\nu}_e}=8\times 10^{51}$\,ergs s$^{-1}$, we find that assuming a constant $Y_{\rm e}$ overestimates $\dot{E}$ by about 7\%. The neutrino reaction rates we use in equations \ref{for1}-\ref{rev2} are approximate and do not take into account all the details as in \cite{Scheck2006}. Given that the physical quantities we measure (see Table \ref{table1}) are not very sensitive to the value of $Y_{\rm e}$ near the PNS surface, the approximate neutrino reaction rates are sufficient for our purposes. These approximations are similar to the approximations we use for the equation of state. This is a trade-off between accuracy and computational complexity. The analytic EOS we use is accurate in the wind temperature range $T\gtrsim 0.5$\,MeV. Most of the characteristics of the wind are determined in these high temperature regions \citep{QW1996}. The wind in this temperature range is composed of relativistic electrons and positrons, non-relativistic baryons and photons. We do not include the formation of $\alpha$ particles or nuclei in the EOS as in \cite{Thompson2001}. We also do not include the effects of strong magnetic fields on the EOS and neutrino emission and absorption rates \citep{Lai1998}. The most important drawback however is that the analytic EOS is not accurate for $T\lesssim 0.5$\,MeV. To assess whether or not the results are sensitive to this approximation, we show results from one simulation run with the Helmholtz EOS \citep{Timmes2000} (see Section \ref{results} for details).

\subsection{Reference frame, initial conditions and boundary conditions}
\label{IC_BC}
In \cite{Prasanna2022}, we perform the simulations in a reference frame rotating with the neutron star. In this work, we choose to perform the simulations in an inertial (lab) frame of reference. Additional source terms have to be considered for the simulations in the rotating frame (see \citealt{Prasanna2022} for details). As the PNS spin period decreases and as the PNS wind approaches a relativistic regime, we have to consider the non-trivial frame transformation of the electric and magnetic fields. We worked out the correct boundary conditions for simulations in the lab frame after our previous paper and we choose to perform the simulations in the lab frame henceforth.

The initial conditions we use are the same as in \cite{Prasanna2022}. We also use the same inner and outer boundary conditions for the radial velocity $v_r$ and $\theta$-velocity $v_{\theta}$, wherein we set $v_r=0$ and $v_{\theta}=0$ at the inner boundary and set outflow boundary conditions at the outer boundary. We emphasize that although $v_r=0$ is not consistent with a steady mass outflow, $v_r=0$ satisfies the axisymmetry condition (that is, we set the $\phi-$derivatives to zero) and works well, and gives consistent results with other choice for the inner velocity boundary condition where we let $v_r$ `float' at the inner boundary by setting the ghost zone value equal to the value at the first active computational zone. Since we perform the simulations in the lab frame in this work, the inner boundary condition for the $\phi$-velocity is $v_{\phi}=r\Omega_{\star} {\rm sin}\,\theta$. $v_{\phi}$ at the outer boundary is set by conserving angular momentum. The temperature at the inner boundary is set by enforcing net zero neutrino heating at the surface of the PNS. The condition of net zero neutrino heating at the PNS surface follows from the kinetic equilibrium established between the neutrino flux and the wind near the surface due to high temperature and density \citep{Burrows1982, QW1996}. Density at the inner boundary depends exponentially on the square of the angular frequency of rotation of the PNS $\Omega_{\star}^2$ as follows (see \citealt{Prasanna2022} for the derivation): 
\begin{equation}\label{densBC}
\begin{split}
    \rho(r,\theta)=\rho_0\exp\left[\frac{GM_{\star}m_{\rm n}}{k_{\rm B}T_0} \left(\frac{1}{r} -\frac{1}{R_{\star}}\right)\right] \exp\left[\frac{m_{\rm n}r^2\Omega_{\star}^2\sin^2\theta }{2k_{\rm B}T_0}\right] \\
    \times \exp \left[\frac{-m_{\rm n} R_{\star}^2\Omega_{\star}^2}{2k_{\rm B}T_0}\right],
    \end{split}
\end{equation}
where we have assumed that the inner boundary is at a constant temperature and that the pressure near the surface is dominated by ideal nucleon pressure. This equation comes from enforcing the axisymmetry condition in the continuity equation (eqn. \ref{eq:continuity}). In equation \ref{densBC}, $m_{\rm n}$ is the average mass of a nucleon and $\rho_0=10^{12}$\,g cm$^{-3}$ is the surface density of the PNS in the non-rotating case. We have tested our 1D non-rotating non-magnetic (NRNM) simulations with base densities $10^{11}-10^{13}$\,g cm$^{-3}$. We find that the mass outflow rate is not very sensitive to the base density over this base density range. We note that eqn. \ref{densBC} assumes a spherical inner boundary which is not a good approximation for millisecond rotation rates. For very rapid rotation with $P_{\star}<5$\,ms, the PNS begins to deform into an ellipsoid. Since we cannot currently include this effect in our simulations, we do not attempt simulations with $P_{\star}<1.5$\,ms (see Section \ref{snapshots}). The electron fraction at the inner and outer boundaries `float', that is, these are set equal to the values of $Y_{\rm e}$ at the first and last active zone respectively. 
 
We set the $\phi$ component of the electric field to zero at the inner edge of the first active computational zone in order to maintain a constant magnetic field at the surface of the PNS with time. Since $v_{\phi}$ is non-zero at the surface in the lab frame, and the electric field $\mathbfit{E}$ of a perfectly conducting fluid is related to the velocity $\mathbfit{v}$ and magnetic field \mathbfit{B} through $\mathbfit{E}=-\mathbfit{v}\times \mathbfit{B}/c$, where $c$ is the speed of light, we do not enforce any condition on $E_{\theta}$ at the surface. Although we do not specify $E_{\theta}$ and $E_{\phi}$ at the surface, the components $B_r$ and $B_{\theta}$ are constant (the constant value depending on $B_0$) in time at the surface by axisymmetry, that is, the $\phi$-derivatives of the electric field components are zero. These follow from $\frac{\partial \mathbfit{B}}{\partial t}=-c\nabla \times \mathbfit{E}$. The component $B_{\phi}$ is automatically specified once we fix $B_r$ and $B_{\theta}$ because $\nabla \cdot \mathbfit{B}=0$. We note that the magnetic field components are constant only at the inner edge of the first active zone and not throughout the first active zone because $E_{\phi}$ is set to zero only at the inner edge. If we set $E_{\phi}$ to zero at the inner edge of the first and second active zone, the magnetic field components then are constant throughout the first active zone. Both of these conditions near the surface result in the measured physical quantities (see Table \ref{table1}) being roughly equal, for instance, we get a less than $1\%$ deviation in simulations with polar magnetic field strength $B_0=10^{15}$\,G, rotation period $P_{\star}=4$\,ms and electron anti-neutrino luminosity $L_{\rm \bar{\nu}_e}=8\times10^{51}$\,ergs s$^{-1}$ at a resolution of $(N_r\,,\,N_{\theta})=(512\,,\,256)$.

\subsection{Resolution of the grid}
\label{res}
We use a logarithmically spaced grid in the radial direction with $N_r$ zones. For rapid rotation, a uniformly spaced grid in the $\theta$ direction is not ideal. If the $\theta$ zones are not closely spaced near the poles, we see effects such as the fluid pressure dropping below the degeneracy pressure (minimum pressure at a given density). Thus, the $\theta$ zones have to be more closely spaced near the poles compared to the zones near the equator. Hence, we use a grid with non-uniform spacing in the $\theta$ direction. We choose the following custom function $g(x_2)$ to define the spacing between the $\theta$ zones:
\begin{equation}
    g(x_2)=(1-w(x_2))\theta_{\rm min} + w(x_2)\theta_{\rm max},
\end{equation}
with $w(x_2)$ defined as
\begin{equation}
    w(x_2)=-1.3x_2^3+1.95x_2^2+0.35x_2,
\end{equation}
where $x_2=i/N_{\theta}$ with $0\leq i \leq N_{\theta}$ is the logical location of the $i^{\rm th}$ $\theta$-zone, $N_{\theta}$ is the total number of $\theta$-zones, $\theta_{\rm min}$ and $\theta_{\rm max}$ are respectively the minimum and maximum values of $\theta$ on the grid. We ensure that the sizes of consecutive $\theta$ zones do not differ by more than $10\%$. With the choice of this function, the zones at the poles are closer by a factor of about 3.7 (the exact factor varies between $3.6-3.8$ depending on the total number of $\theta$ zones) at the pole than at the equator. For example, with a total of 512 $\theta$ zones, the zones at the poles are separated by $0.124^{\circ}$ while the zones at the equator are separated by $0.466^{\circ}$.      

\section{Results}
\label{results}
\begin{table*}
	\caption{Wind properties at $L_{\rm \bar{\nu}_e}=8\times10^{51}$\,ergs s$^{-1}$ for a $1.4$\,M$_{\odot}$ PNS. All the velocities are in the units of speed of light $c$.}
	\label{table1}
      \begin{threeparttable}
	\begin{tabular}[width=\textwidth]{@{}cccccccccccccc@{}} 
		\hline
		 $B_0$ & $P_{\star}$ &$\dot{E}$& $\tau_{\rm J}$& $\dot{J}$ & $\dot{M}$ & $R_{\rm A}^{\rm sph}$ &$\langle R_{\rm A} \rangle$ & $\langle R_{\rm son} \rangle$ &$\Delta t_{\rm p}$ & Max $\rm c_s$ & Max $\rm v_A$ & Max FMS &$\langle v_{\rm A}\rangle_{\dot{M},R_{\rm A}}^{\&}$\\
		(G) & (ms)&(ergs s$^{-1}$) & (s) & (g cm$^{2}$\,s$^{-2}$) &($\rm M_{\odot}$ s$^{-1}$) & (km) &(km) & (km) &(ms) &($c$)&($c$)&($c$)&($c$)\\
		\hline\hline \\
		 $3 \times 10^{15}$ & 4$^{\ast}$ & $2.32\times10^{50}$ & 16.7 & $1.63\times10^{47}$ & $3.79\times10^{-4}$ &  117 & 972 & 84 & 26 & 0.55 & 1.05 & 1.05 &0.54  \\
    & $8^{*}$  & $7.80\times10^{49}$ & 12.7 & $1.08\times10^{47}$ & $2.69\times10^{-4}$ & 160  & 1170 & 137 & 34 & 0.50 & 1.22 & 1.23 & 0.38  \\
    & $8^{\$}$  & $7.63\times10^{49}$ & 12.8 & $1.07\times10^{47}$ & $2.74\times10^{-4}$ & 158  & 663 & 163 & 35 & 0.59 & 0.84 & 0.85 & 0.38  \\
    & 16  & $2.94\times10^{49}$ & 8.8 & $7.59\times10^{46}$ & $2.44\times10^{-4}$ & 200 & 1452 & 220 & 35 & 0.35 & 0.62 & 0.62 & 0.23   \\
    & 32  & $1.52\times10^{49}$ & 5.6 & $5.74\times10^{46}$ & $2.41\times10^{-4}$ & 247  & 1289 & 327 & 39 & 0.32 & 0.55 & 0.55 &0.16  \\ \\
    
    $2 \times 10^{15}$ & 2$^{\ast}$  & $6.22\times10^{50}$ & 21.1 & $2.43\times10^{47}$ & $1.03\times10^{-3}$ &  61  & 1050 & 170 & 14 & 2.49 & 4.73 & 4.73 &0.48   \\ \\

    $10^{15}$ & 1.5$^{\ast}$  & $4.21\times10^{50}$ & 34.1 & $1.98\times10^{47}$ & $2.04\times10^{-3}$&  34   & 327 & 71 & - & 0.30 & 0.92 & 0.92&0.30   \\
    & 2  & $2.49\times10^{50}$& 41.3 & $1.23\times10^{47}$ & $9.03\times10^{-4}$ &  47   & 396 & 82 & - & 0.18 & 0.76 & 0.76 & 0.30   \\
    & 4  & $8.55\times10^{49}$ & 32.1 & $7.91\times10^{46}$ & $4.82\times10^{-4}$&  72  & 431 & 102 & - & 0.16 & 0.34 & 0.34 &0.26   \\
    & 8  & $3.10\times10^{49}$ & 22.6 & $5.60\times10^{46}$ & $3.45\times10^{-4}$& 102   & 481 & 162 & - & 0.16 & 0.29 & 0.29&0.18   \\
    & $8^{\#}$  & $3.14\times10^{49}$ & 22.5 & $5.63\times10^{46}$ & $3.45\times10^{-4}$&  102  & 514 & 161 & - & 0.16 & 0.38 & 0.38&0.18   \\
    & 16  & $1.23\times10^{49}$ & 14.7 & $4.30\times10^{46}$ & $3.19\times10^{-4}$& 131   & 512 & 255 & - & 0.16 & 0.18 & 0.19 & 0.11   \\ 
    & 32  & $5.44\times10^{48}$ & 10.2 & $3.10\times10^{46}$ & $3.02\times10^{-4}$& 162   & 511 & 357 & - & 0.17 & 0.19 & 0.19 & 0.08  \\ \\

    $3\times10^{14}$& 1.5  & $4.84\times10^{49}$ & 173.5 & $3.98\times10^{46}$ & $8.77\times10^{-4}$& 23   & 532 & 137 & - & 0.17 & 0.60 & 0.60 &0.17  \\
    & 2  & $3.32\times10^{49}$ & 134.9 & $3.89\times10^{46}$ & $7.39\times10^{-4}$& 29   & 295 & 90 & - & 0.18 & 0.58 & 0.58 & 0.14  \\
    & 4  & $1.53\times10^{49}$ & 100.7 & $2.52\times10^{46}$ & $4.21\times10^{-4}$& 44   & 226 & 187 & - & 0.16 & 0.16 & 0.17 &0.12   \\
    & 8  & $6.17\times10^{48}$ & 78.7 & $1.61\times10^{46}$ & $3.34\times10^{-4}$& 56   & 257 & 282 & - & 0.16 & 0.12 & 0.16&0.08   \\
    & 16  & $3.24\times10^{48}$ & 58.6 & $1.08\times10^{46}$ & $3.07\times10^{-4}$& 67   & 248 & 362 & - & 0.16 & 0.08 & 0.16 &0.05   \\ 
    & 32  & $2.19\times10^{48}$ & 49.2 & $6.44\times10^{45}$ & $2.99\times10^{-4}$&  74  & 193 & 432 & - & 0.16 & 0.06 & 0.16 &0.04   \\ \\

    $10^{14}$& 1.5  & $1.75\times10^{49}$ & 274.1 & $2.49\times10^{46}$ & $7.36\times10^{-4}$& 20   & 267 & 152 & - & 0.22 & 0.80 & 0.80 &0.11   \\
    & 2  & $6.18\times10^{48}$ & 374.2 & $1.37\times10^{46}$ & $5.04\times10^{-4}$& 21  & 245 & 207 & - & 0.32 & 0.54 & 0.54 & 0.09  \\
    & 4  & $4.40\times10^{48}$ & 223.4 & $1.13\times10^{46}$ & $3.75\times10^{-4}$& 31  & 205 & 312 & - & 0.16 & 0.16 & 0.17 &0.07   \\
    & 8  & $2.42\times10^{48}$ & 237.6 & $5.33\times10^{45}$ & $3.14\times10^{-4}$&  33  & 154 & 391 & - & 0.16 & 0.06 & 0.16 & 0.04   \\
    & 16  & $1.92\times10^{48}$ & 224.8 & $2.82\times10^{45}$ & $3.01\times10^{-4}$& 35  & 105 & 442 & - & 0.16 & 0.04 & 0.16 &0.02   \\
    & 32  & $1.76\times10^{48}$ & 216.7 & $1.46\times10^{45}$ & $2.97\times10^{-4}$& 35  & 78 & 465 & - & 0.16 & 0.03 & 0.16 & 0.02   \\ \\

   $10^{13}$ & 1$^{\ast}$  & $3.54\times10^{48}$ & 945.5 & $1.07\times10^{46}$ & $6.18\times10^{-4}$&  12  & 138 & 294 & - & 0.30 & 0.52 & 0.52 &0.06  \\
   & 1.5  & $2.47\times10^{48}$ & 1121.7 & $6.03\times10^{45}$ & $4.34\times10^{-4}$& 13   & 96 & 353 & - & 0.18 & 0.17 & 0.18 &0.05   \\
   & 2  & $2.13\times10^{48}$ & 1274.6 & $3.98\times10^{45}$ & $3.69\times10^{-4}$& 13  & 87 & 404 & - & 0.17 & 0.09 & 0.17 &0.04  \\
   & 4  & $1.82\times10^{48}$ & 1375.1 & $1.84\times10^{45}$ & $3.14\times10^{-4}$&  14  & 64 & 461 & - & 0.16 & 0.05 & 0.16 &0.02  \\
   & 8  & $1.73\times10^{48}$ & 1405.2 & $9.03\times10^{44}$ & $3.01\times10^{-4}$& 14   & 31 & 472 & - & 0.16 & 0.03 & 0.16 &0.01  \\
   & 16  & $1.70\times10^{48}$ & 1529.0 & $4.14\times10^{44}$ & $2.97\times10^{-4}$& 13    & 16 & 475 & - & 0.16 & 0.008 & 0.16 & 0.005  \\
   & 32  & $1.69\times10^{48}$ & 1597.4 & $1.98\times10^{44}$ & $2.96\times10^{-4}$&13    & 18 & 478 & - & 0.16 & 0.004 & 0.16 &0.003  \\

 \hline
	   \end{tabular}
    \begin{tablenotes}
       \item[\$] Simulation with outer grid boundary at 3000\,km. Rest of the simulations are with outer grid boundary at 10000\,km (see Section \ref{snapshots} for details).
        \item[\#] Simulation with Helmholtz EOS. Rest of the simulations are with analytic EOS from \citealt{QW1996}.
        \item[\&] Maximum value of $\dot{M}$ averaged Alfv\'en speed at the Alfv\'en surface during the simulation. 
        \item[*] These results maybe \textit{unreliable}, see Section \ref{results} for details.
    \end{tablenotes}
      \end{threeparttable}
\end{table*}
\subsection{Diagnostic quantities}
The mass outflow rate $\dot{M}$, angular momentum flux $\dot{J}$, spindown timescale $\tau_{\rm J}$, and energy outflow rate $\dot{E}$ are the principal physical quantities we measure. The mass outflow rate is given by
\begin{equation}
\label{Mdot}
    \dot{M} \left(r \right)= \oint_S r^2 \rho v_r d\Omega.
\end{equation}
The $z$-component (along the axis of rotation of the PNS) of the angular momentum flux is given by the following integral over a closed spherical surface \citep{Vidotto2014}:
\begin{equation}
\label{Jdot}
    \dot{J}\left(r \right)= \oint_S \left[-\frac{B_rB_{\phi}r\sin \theta}{4\pi}+\rho v_r v_{\phi}r\sin \theta\right]r^2 d\Omega.
\end{equation}
The total angular momentum of the star is roughly $J=\frac{2}{5}MR_{\star}^2\Omega_{\star}$. We define the spindown time of the PNS as:
\begin{equation}
\label{tauj}
    \tau_{\rm J}=\frac{J}{\dot{J}}.
\end{equation}
The energy flux is given by the following surface integral (we generalize the definition in \citealt{Metzger2007} to two and three dimensions):
\begin{align}
\label{Edot}
    \dot{E}\left(r \right)&= \oint_S r^2 \rho v_r\left[\frac{1}{2}\left(v_r^2+v_{\theta}^2+v_{\phi}^2\right) -\frac{rB_rB_{\phi}\Omega_{\star}\sin\theta}{4 \pi \rho v_r}\right.  \\  \nonumber &\qquad -\left.\frac{GM_{\star}}{r} +e+\frac{P}{\rho}\right] d\Omega,
\end{align}
where $e$ is the specific internal energy of the outflow. 

To define the magnetosonic surfaces, we use the total velocity. At the Alfv\'en surface we have,
\begin{equation}
\label{s_alph}
    v_r^2+v_{\theta}^2+v_{\phi}^2=\frac{B_r^2+B_{\theta}^2+B_{\phi}^2}{4\pi \rho}.
\end{equation}
At the adiabatic sonic surface, the magnitude of velocity is equal to the adiabatic sound speed $c_{\rm s}$ and at the fast magnetosonic surface we have,
\begin{equation}
    v_r^2+v_{\theta}^2+v_{\phi}^2=\frac{1}{2}\left(v_{\rm A}^2+c_{\rm s}^2+ \sqrt{\left(v_{\rm A}^2+c_{\rm s}^2\right)^2-4v_{\rm A}^2c_{\rm s}^2 {\rm cos}^2 \Theta}\right),
\end{equation}
where $\Theta$ is the angle between the magnetic field and the direction of wave propagation. We assume that the magnetosonic waves propagate radially and thus, we have ${\rm \cos} \ \Theta=\mathbfit{B}\cdot \hat{\mathbfit{r}}/B$, where $B$ is the magnitude of the magnetic field. 
Since we focus on rapid rotation in this paper, it is important to include the $\phi$-components of velocity and magnetic fields while computing the magnetosonic speeds. Moreover, we compute the magnetosonic speeds to track the transition from non-relativistic to the relativistic regime. Hence, including the $\phi$-components is important so that we do not underestimate the magnetosonic speeds. The spherical Alfv\'en radius can be obtained from $\dot{J}$ and $\dot{M}$ \citep{Raives2023}:
\begin{equation}
\label{Ra_sph}
 R_{\rm A}^{\rm sph}=\sqrt{\frac{\dot{J}}{\dot{M}\Omega_{\star}}}.   
\end{equation}
$R_{\rm A}^{\rm sph}$ is a measure of the radius to which the wind effectively co-rotates with the PNS (see Table \ref{table1} and Section \ref{snapshots} for details).\\

\begin{figure*}
\centering{}
\includegraphics[width=\textwidth]{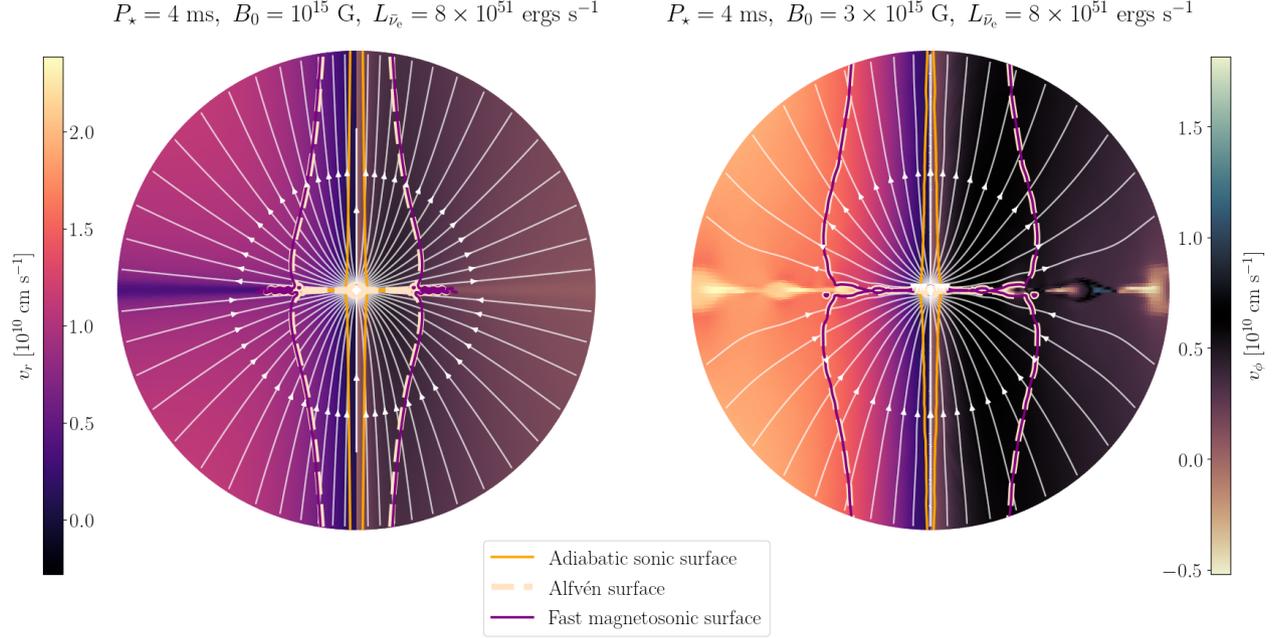}
\caption{2D map of $v_r$ (left half of each panel) and $v_{\phi}$ (right half of each panel) at a fixed rotation period of $4$\,ms and electron type anti-neutrino luminosity of $8\times 10^{51}$\,ergs s$^{-1}$. The small white circle at the center of each panel is the PNS with a radius of 12\,km. The rotation and magnetic axes of the PNS are along the vertical and the outer boundary in the figure is at $500$\,km. The outer boundary is at a radius of 10000\,km in the simulation. The white lines with arrows are the magnetic field lines. Plasmoids begin to occur as $B_0$ increases and the magnetic energy density at the equator exceeds the fluid pressure (see \citealt{Prasanna2022} for details). The Alfv\'en surface moves outwards as $B_0$ increases.}
\label{vrvphi_B0}
\end{figure*}

\begin{figure*}
\centering{}
\includegraphics[width=\textwidth]{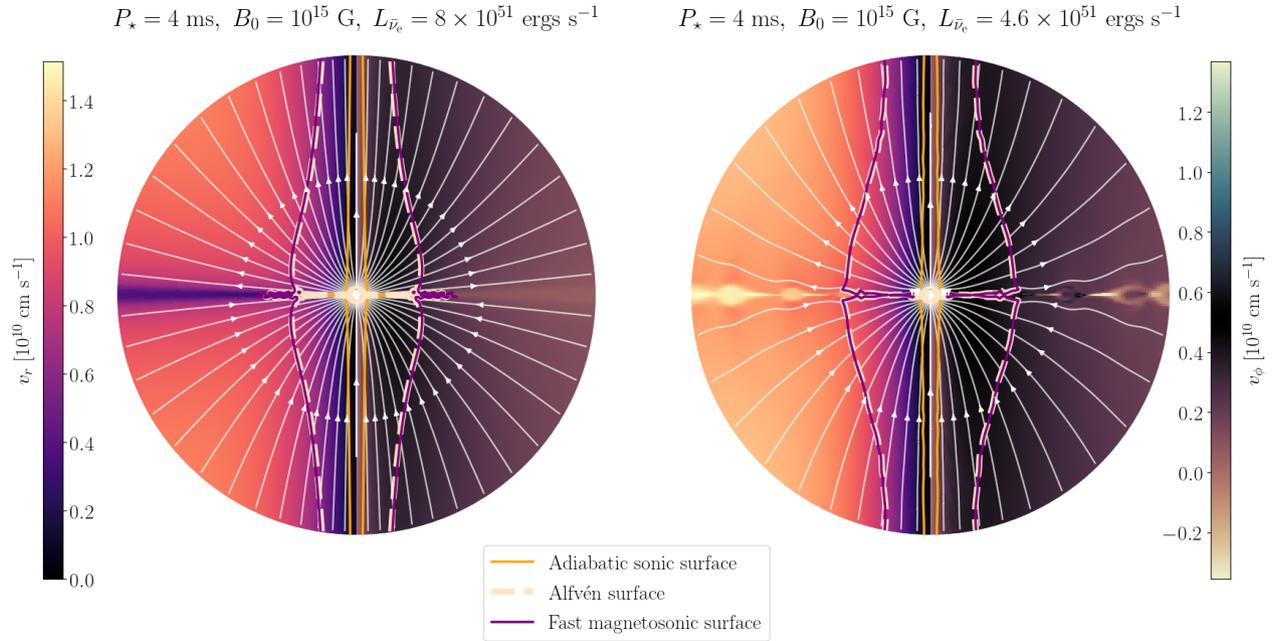}
\caption{Same as Figure \ref{vrvphi_B0}, but at a fixed value of $B_0$ and varying neutrino luminosity. As the neutrino luminosity decreases, the fluid pressure decreases and plasmoids develop when the fluid pressure at the equator drops below the magnetic energy density. The Alfv\'en surface moves outwards as the neutrino luminosity decreases.}
\label{vrvphi_Lnubar}
\end{figure*}

\begin{figure*}
\centering{}
\includegraphics[width=\textwidth]{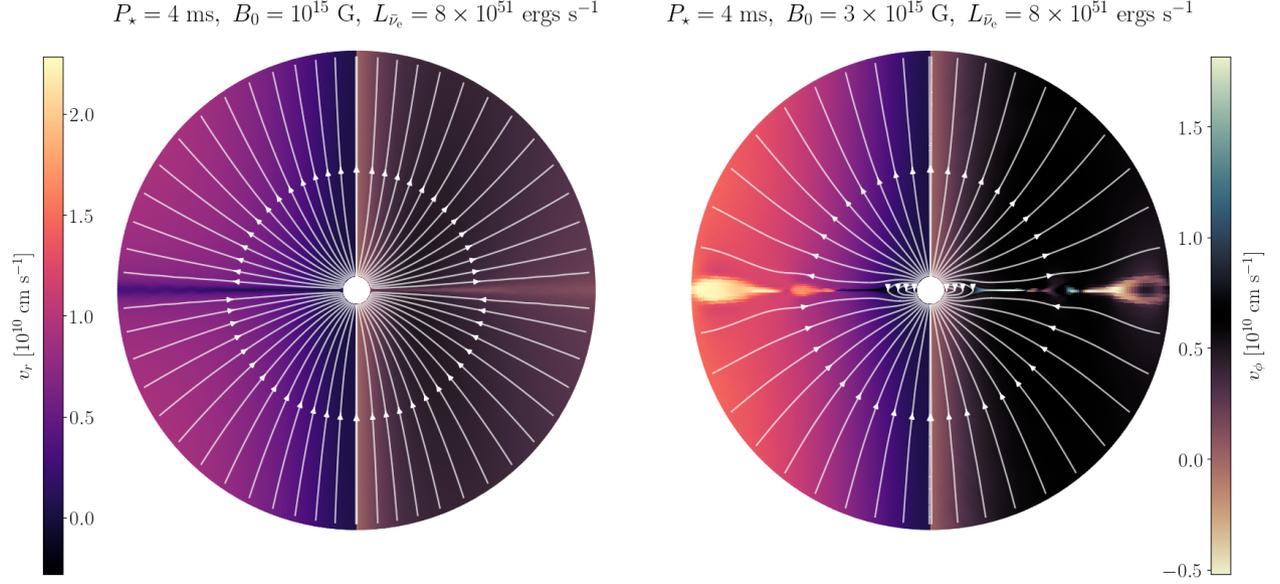}
\caption{Zoomed in version of Figure \ref{vrvphi_B0} with the outer boundary in the figure at a radius of 200\,km. The white circle at the center of each panel is the PNS with a radius of 12\,km.}
\label{vrvphi_B0_zoom}
\end{figure*}

\begin{figure*}
\centering{}
\includegraphics[width=\textwidth]{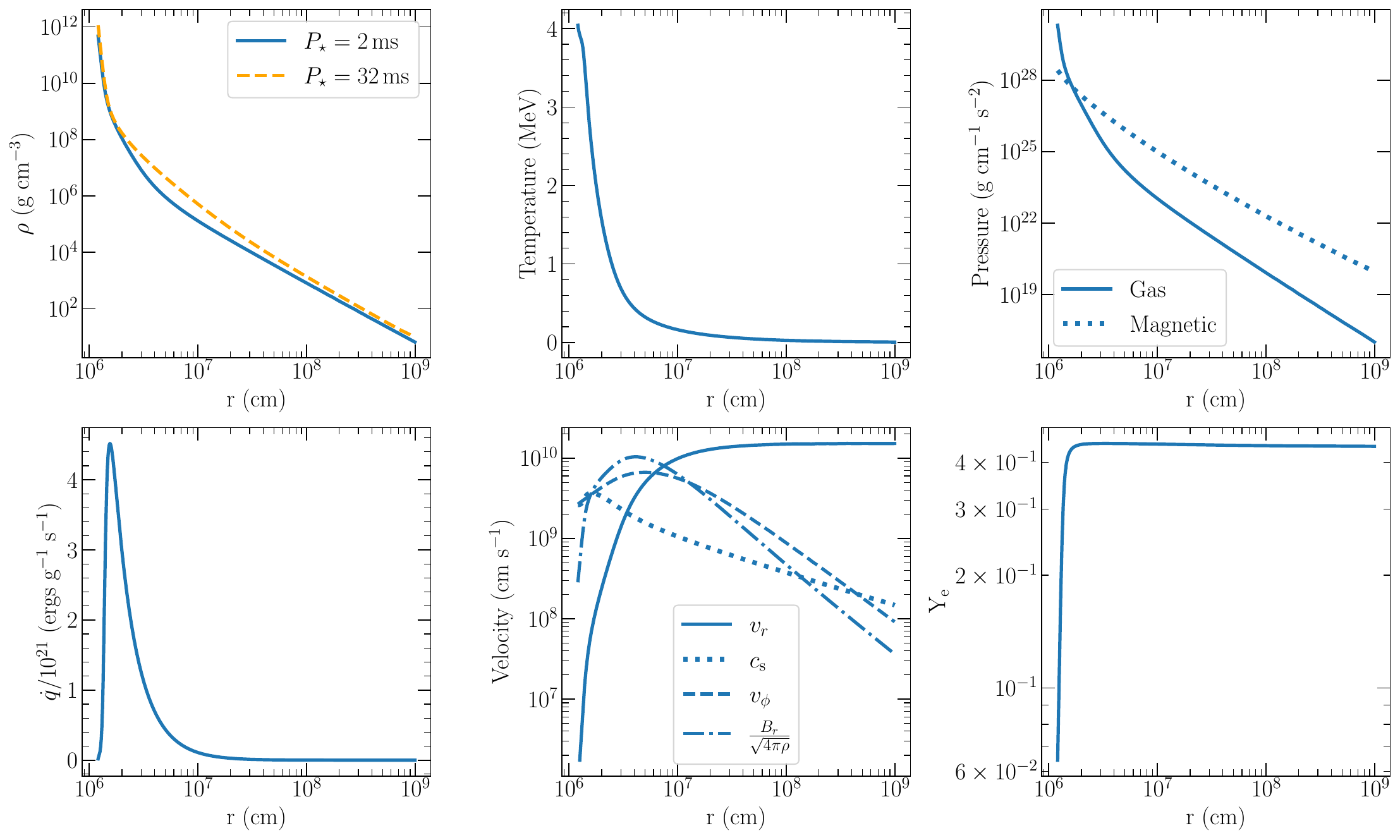}
\caption{1D profiles at $\theta=45 ^{\circ}$ as a function of radius in the simulation with $B_0=10^{15}$\,G, $P_{\star}=2$\,ms and $L_{\rm \bar{\nu}_e}=8\times10^{51}$\,ergs s$^{-1}$. The simulation reaches a steady state for these parameters. The panels show density, temperature, pressure, specific neutrino heating/cooling rate $\dot{q}$, velocity profiles and the electron fraction $Y_{\rm e}$. The first (density) panel shows the density profile at $P_{\star}=32$\,ms too for comparison.}
\label{1d_th45}
\end{figure*}

\begin{figure*}
\centering{}
\includegraphics[width=\textwidth]{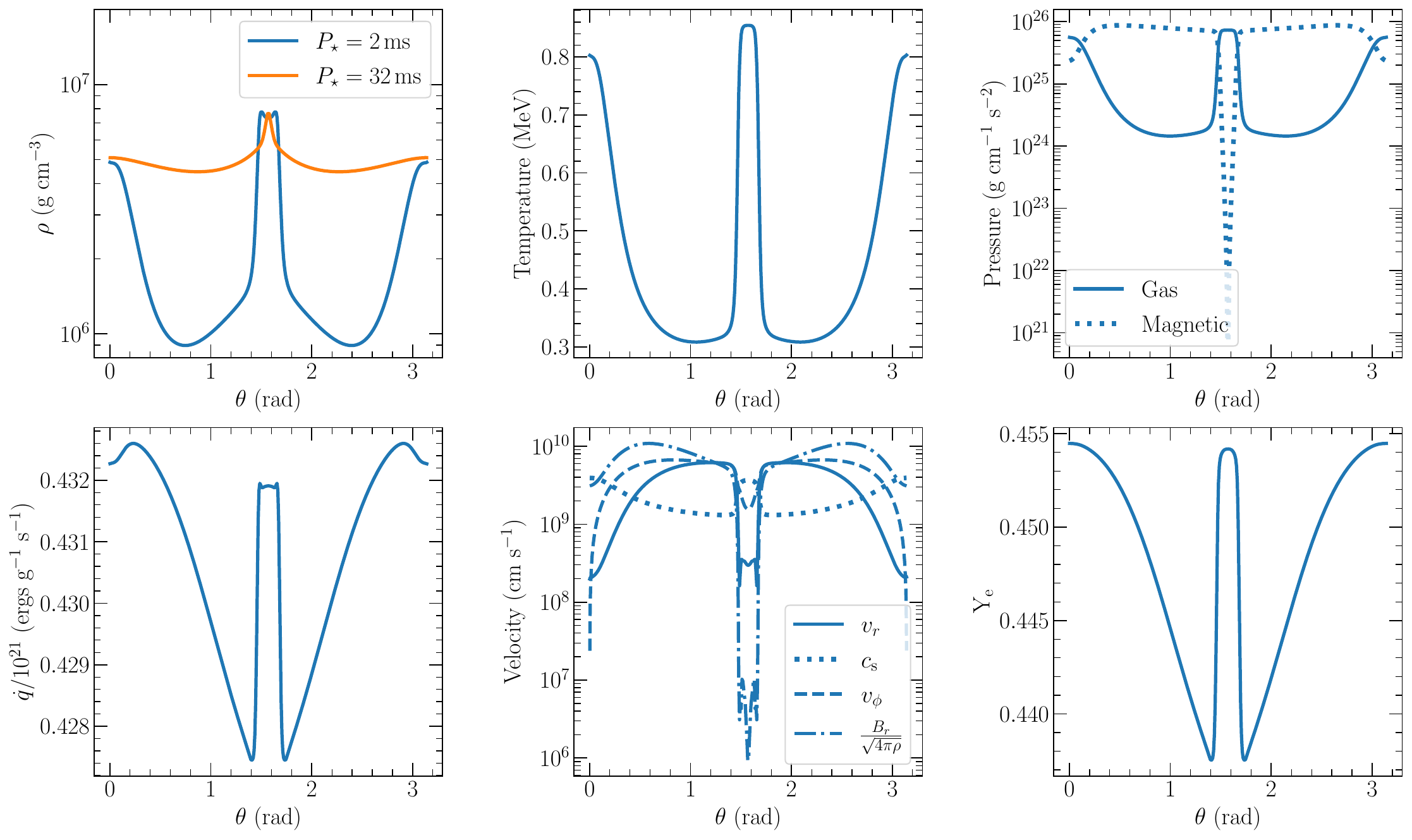}
\caption{The various physical quantities have been plotted as a function of polar angle $\theta$ at a fixed radius of $50$\,km for the same simulation as in Figure \ref{1d_th45}.}
\label{1d_r50}
\end{figure*}

\begin{figure}
\centering{}
\includegraphics[width=\linewidth]{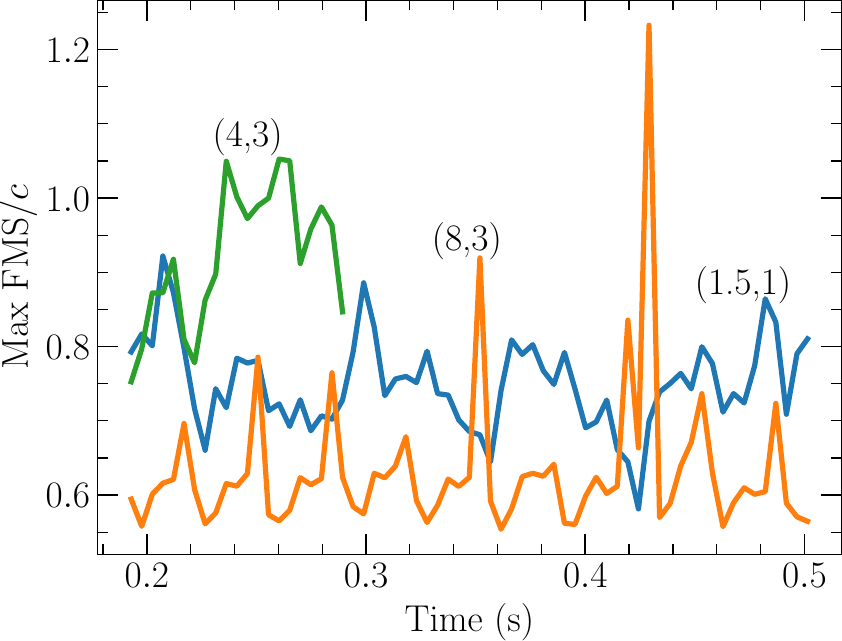}
\caption{Maximum value of the Fast Magnetosonic Speed (FMS) at each time instant over the course of the simulation. The numbers in the parentheses are PNS spin period in milliseconds and polar magnetic field strength $B_0$ in units of $10^{15}$\,G. Due to computational complexity, we run the simulation with $P_{\star}=4$\,ms and $B_0=10^{15}$\,G for only 0.3\,s. We show the plot starting from $\sim 0.2$\,s so as to not include the initial transient.}
\label{max_FMS}
\end{figure}

\subsection{Time-steady snapshots}
\label{snapshots}
We first present results of snapshots from simulations at a fixed neutrino luminosity and PNS spin period. Figure \ref{vrvphi_B0} shows 2D maps of $v_r$ and $v_{\phi}$ at a PNS spin period of 4\,ms and electron anti-neutrino luminosity of $L_{\rm \bar{\nu}_e}=8\times 10^{51}$\,ergs s$^{-1}$ for different values of polar magnetic field strength $B_0$. As $B_0$ increases, the Alfv\'en surface moves away from the PNS. As $B_0$ increases and the magnetic energy density at the equator exceeds the fluid pressure, plasmoids begin to develop (see \citealt{Thompson2003}, \citealt{Thompson2018}, \citealt{Prasanna2022} for details about plasmoids). Figure \ref{vrvphi_Lnubar} shows the 2D maps of $v_r$ and $v_{\phi}$ at $B_0=10^{15}$\,G and $P_{\star}=4$\,ms for different values of $L_{\rm \bar{\nu}_e}$. As $L_{\rm \bar{\nu}_e}$ decreases, the fluid pressure decreases and plasmoids develop when the fluid pressure drops below the magnetic energy density at the equator. The maximum values of $v_r$ and $v_{\phi}$ in the figures are in excess of $10^{10}$\,cm s$^{-1}$ compared to $\sim 10^{9}$\,cm s$^{-1}$ for a PNS rotating slowly ($P_{\star}\gtrsim 50$\,ms, see Figures 1 and 2 in \citealt{Prasanna2022}). The enhancement in the values $v_r$ and $v_{\phi}$ is due to the larger angular velocity of rotation of the PNS and magneto-centrifugal acceleration. Figure \ref{vrvphi_B0_zoom} shows a zoomed in version of Figure \ref{vrvphi_B0} with the central PNS and the structure of the magnetosphere.   

We present results from a calculation at a fixed electron anti-neutrino luminosity of $L_{\rm \bar{\nu}_e}=8\times 10^{51}$\,ergs s$^{-1}$ that is representative of the first few seconds after the explosion as the PNS begins to cool. We assume mean neutrino energies of $\langle\epsilon_{\rm \bar{\nu}_e}\rangle=14$\,MeV and $\langle\epsilon_{\rm \nu_e}\rangle=11$\,MeV. Figure \ref{1d_th45} shows the line plots at $\theta=45^{\circ}$ slice in a simulation with $P_{\star}=2$\,ms and polar magnetic field strength $B_0=10^{15}$\,G. The dotted line in the density panel shows the density profile at $P_{\star}=32$\,ms and $B_0=10^{15}$\,G for comparison. As the density at the surface depends exponentially on $\Omega_{\star}^2$ and $\sin^2 \theta$ (see eqn. \ref{densBC}), noticeable differences in base density as a function of $\theta$ occur for $P_{\star}\leq 3$\,ms. The last panel in Figure \ref{1d_th45} shows the profile of electron fraction $Y_{\rm e}$. The electron fraction increases rapidly near the PNS surface and achieves asymptotic value within $1-2$ PNS radii. Figure \ref{1d_r50} shows the physical quantities as a function of polar angle $\theta$ at a radius of 50\,km for a PNS with $B_0=10^{15}$\,G and $P_{\star}=2$\,ms. The asymmetry caused by rapid rotation is evident in all the panels. In the density panel, we show the density profile at $P_{\star}=32$\,ms for comparison. As expected, we find that as the PNS spin period increases, the profiles approach spherical symmetry.

Table \ref{table1} presents the physical quantities for a 1.4\,M$_{\odot}$ PNS at $L_{\rm \bar{\nu}_e}=8\times 10^{51}$\,ergs s$^{-1}$, namely the asymptotic energy outflow rate $\dot{E}$, spindown time $\tau_{\rm J}$, angular momentum flux $\dot{J}$, mass outflow rate $\dot{M}$, the spherical Alfv\'en radius $R_{\rm A}^{\rm sph}$, angle averaged Alfv\'en radius $R_{\rm A}$, angle averaged adiabatic sonic radius $R_{\rm son}$, time between plasmoids $\Delta t_{\rm p}$, maximum adiabatic sonic speed $c_{\rm s}$, maximum Alfv\'en speed $v_{\rm A}$, maximum fast magnetosonic speed (FMS) and maximum average value of the Alfv\'en speed at the Alfv\'en surface weighted by $\dot{M}$. By maximum, we mean the maximum value of the magnetosonic speeds over the time of the simulation.  The physical quantities (except the maximum values of the magnetosonic speeds) have been averaged over time to account for plasmoids. We track the maximum magnetosonic speeds in order to ensure they are less than the speed of light, as our simulations use non-relativistic physics. We measure $\dot{J}$ and $\dot{M}$ at a radius of 50\,km and asymptotic $\dot{E}$ at a radius of 1000\,km. We note that the time-averaged value of $\dot{J}$ and $\dot{M}$ is negligibly dependent on the radius of measurement. The time-averaged value of asymptotic $\dot{E}$ is also negligibly dependent on the radius of measurement as long as it is measured at radii well past the peak of net neutrino heating rate (see the first panel on the second row of Figure \ref{1d_th45}). We compute the angle-averaged Alfv\'en radius $\langle R_{\rm A}\rangle$ and adiabatic sonic radius $\langle R_{\rm son}\rangle$ from the outputs of the simulations. As shown in Figures \ref{vrvphi_B0} and \ref{vrvphi_Lnubar}, the magnetosonic surfaces approach a cylindrical shape as the PNS rotates faster and as the polar magnetic field strength $B_0$ increases. $\langle R_{\rm A}\rangle$ is not a measure of the radius to which effective co-rotation of the PNS wind is enforced because the energy outflow rate $\dot{E}$ is small near the poles. On the other hand, $R_{\rm A}^{\rm sph}$ (see eqn. \ref{Ra_sph}) which is related to the specific angular momentum carried by the PNS wind is a direct measure of the radius to which effective co-rotation of the wind is enforced. We note that $\langle R_{\rm A}\rangle$ is much larger than $R_{\rm A}^{\rm sph}$ at large values of $B_0$ and PNS rotation rate (see Table \ref{table1}) due to highly cylindrical nature of the magnetosonic surfaces. 

The asterisk on some calculations in Table \ref{table1} is to emphasize that those results are from possibly \textit{unreliable} simulations. Since we use non-relativistic physics, we deem a simulation to be  ``possibly \textit{unreliable}" when the magnetosonic speeds exceed the speed of light $c$. Although the maximum magnetosonic speeds are larger than $c$ in some simulations, the $\dot{M}$ weighted average speeds at the Alfv\'en surface are below $c$. A related quantity used in literature is the magnetization $\sigma$ (see \citealt{Metzger2011} for example).  The $\dot{M}$ weighted speed is likely a better indicator because it is a measure of the magnetosonic speed in the regions of the computational domain where the mass flux is the largest. Since the $\dot{M}$ weighted average speeds at the Alfv\'en surface are below $c$, we can claim with some confidence that the calculations in which the maximum magnetosonic speeds exceed $c$ are in fact \textit{reliable}. When the maximum of the magnetosonic speed over the time of a simulation is larger than $c$, we emphasize that this does not occur throughout the computational domain and at all times, except for the simulation with $P_{\star}=2$\,ms and $B_0=2\times 10^{15}$\,G. Although the maximum magnetosonic speeds for $P_{\star}=2$\,ms and $B_0=2\times10^{15}$\,G are larger than $c$ at all times, we present this result only to get an estimate. We emphasize that this result is not reliable. In the few other simulations with plasmoids with $B_0= 3\times 10^{15}$\,G and $P_{\star}= 4$\,ms and 8\,ms, the maximum magnetosonic speeds exceed $c$ (see Table \ref{table1}). This happens only in small pockets in the plasmoids near the equator and in small pockets near polar angle $\theta\sim 45^{\circ}$. Figure \ref{max_FMS} shows the maximum fast magnetosonic speed (FMS) at each time instant over the course of the simulation for the simulations in which the maximum FMS approaches or exceeds $c$ (we have not shown $P_{\star}=2$\,ms and $B_0=2\times 10^{15}$\,G because the maximum FMS in this simulation is larger than $3c$ at all times). The maximum FMS as a function of time is mostly smaller than $c$ except a few spikes that approach or exceed $c$. It is reasonable to assume that these small pockets of large magnetosonic speeds that last for a very short period of time do not significantly affect the time-averaged results. 

Another region where the magnetosonic speeds exceed $c$ is at the poles close to the outer boundary. These possibly occur due to insufficient resolution at the poles for very rapid rotation with $P_{\star}\leq 4$\,ms. Since the energy outflow rate $\dot{E}$ is negligible near the poles, it is reasonable to expect that these regions do not affect the results significantly. However, we emphasize that relativistic calculations are needed to fully assess these results in the rapidly rotating and highly-magnetic limits as the magnetosonic speeds approach $c$. With the outer boundary at 10000\,km, we find that the highly cylindrical magnetosonic surfaces cannot be captured on the computational domain near the poles during plasmoid eruptions. To test if this affects the results, we have run one simulation at $P_{\star}=8$\,ms and $B_0=3\times 10^{15}$\,G with the outer boundary at 3000\,km (see Table \ref{table1}) and we find that the results from this simulation are close to the results from the simulation with outer boundary at 10000\,km. 

Finally as a test of the dependence of our results on the EOS, we show one result from a simulation with the Helmholtz EOS at $B_0=10^{15}$\,G and $P_{\star}=8$\,ms (see Table \ref{table1}). We find that the results obtained using the Helmholtz EOS are very close to the results obtained using the approximate analytic EOS, which shows that the simplifying assumptions made in order to lower the computational time do not significantly affect the results.  

\begin{figure}
\centering{}
\includegraphics[width=\linewidth]{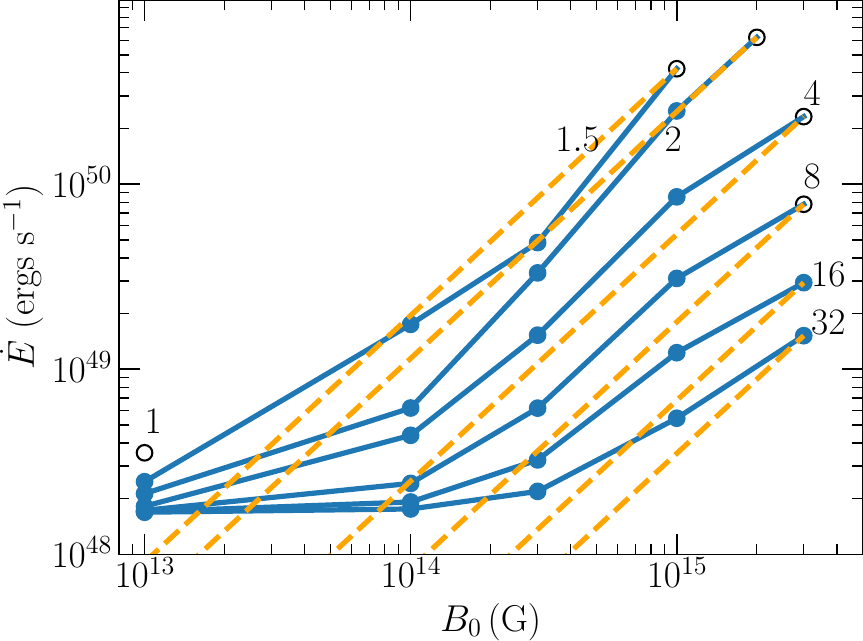}
\caption{Energy outflow rate shown by solid blue lines as a function of polar magnetic field strength at $L_{\rm \bar{\nu}_e}=8\times 10^{51}$\,ergs s$^{-1}$ for a 1.4\,M$_{\odot}$ PNS. The numbers next to the lines represent the PNS spin period $P_{\star}$ in milliseconds. The circles represent actual data points from Table \ref{table1}, which are connected by continuous lines. The unfilled black circles represent results from simulations in which either the maximum Alfv\'en speed exceeds the speed of light (the $\dot{M}$ weighted average Alfv\'en speed in these calculations is below $c$, see Section \ref{results} for details) or the PNS is a 1\,ms rotator with an uncertain density inner boundary condition (see Section \ref{results} for details). The dashed orange lines show the $\dot{E} \propto B_0^{4/3}$ scaling starting from the highest value of $B_0$ for each spin period $P_{\star}$.}
\label{edot_B_P}
\end{figure}

\begin{figure}
\centering{}
\includegraphics[width=\linewidth]{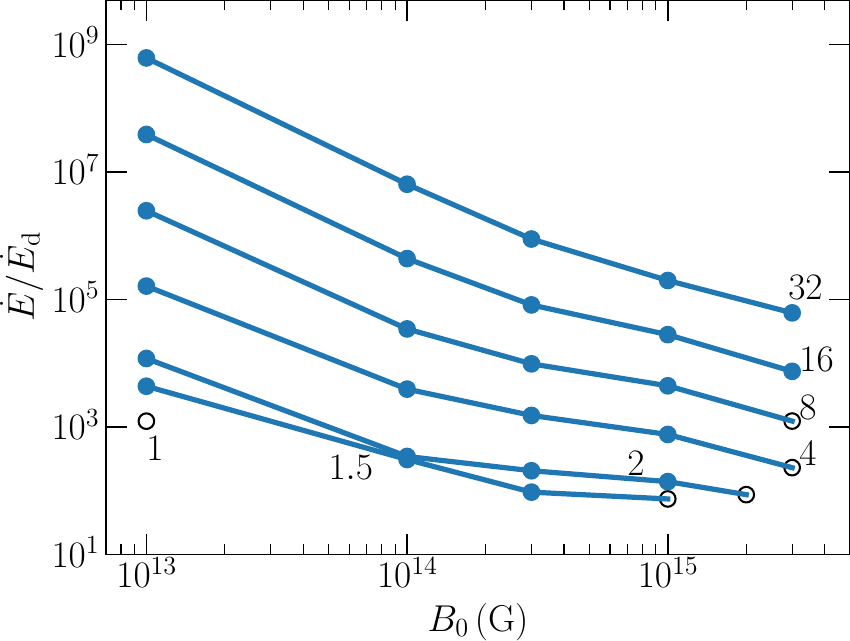}
\caption{Ratio of values of $\dot{E}$ from Figure \ref{edot_B_P} to the energy outflow rate $\dot{E}_{\rm d}$ predicted by vacuum dipole spindown for an orthogonal rotator. The numbers next to the lines represent the PNS spin period in milliseconds. Refer to the caption of Figure \ref{edot_B_P} for the meaning of filled and unfilled circles.}
\label{edot_dip_rat}
\end{figure}

\begin{figure}
\centering{}
\includegraphics[width=\linewidth]{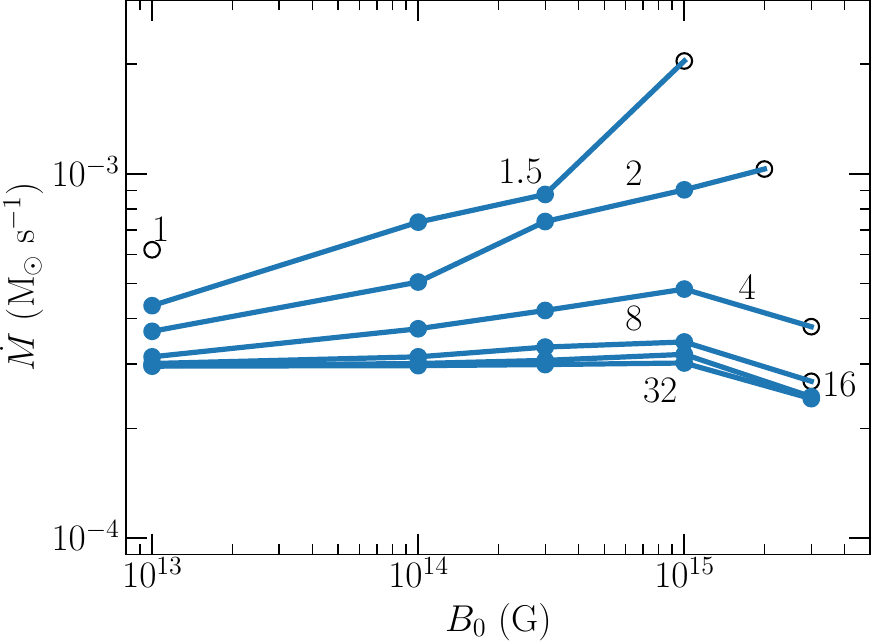}
\caption{Mass outflow rate as a function of polar magnetic field strength and PNS spin period $P_{\star}$ in milliseconds at $L_{\rm \bar{\nu}_e}=8\times 10^{51}$\,ergs s$^{-1}$ for a 1.4\,M$_{\odot}$ PNS. The numbers next to the lines represent the PNS spin period in milliseconds. Refer to the caption of Figure \ref{edot_B_P} for the meaning of filled and unfilled circles.}
\label{mdot_B_P}
\end{figure}

\begin{figure}
\centering{}
\includegraphics[width=\linewidth]{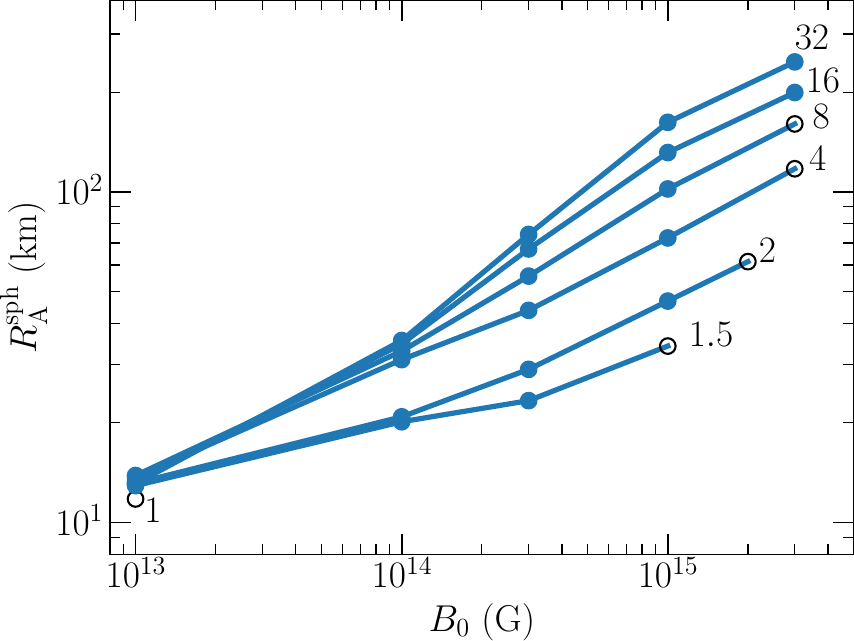}
\caption{Spherical Alfv\'en radius as a function of polar magnetic field strength and PNS spin period $P_{\star}$ in milliseconds at $L_{\rm \bar{\nu}_e}=8\times 10^{51}$\,ergs s$^{-1}$ for a 1.4\,M$_{\odot}$ PNS. Refer to the caption of Figure \ref{edot_B_P} for the meaning of filled and unfilled circles.}
\label{Rasph_B_P}
\end{figure}

\begin{figure}
\centering{}
\includegraphics[width=\linewidth]{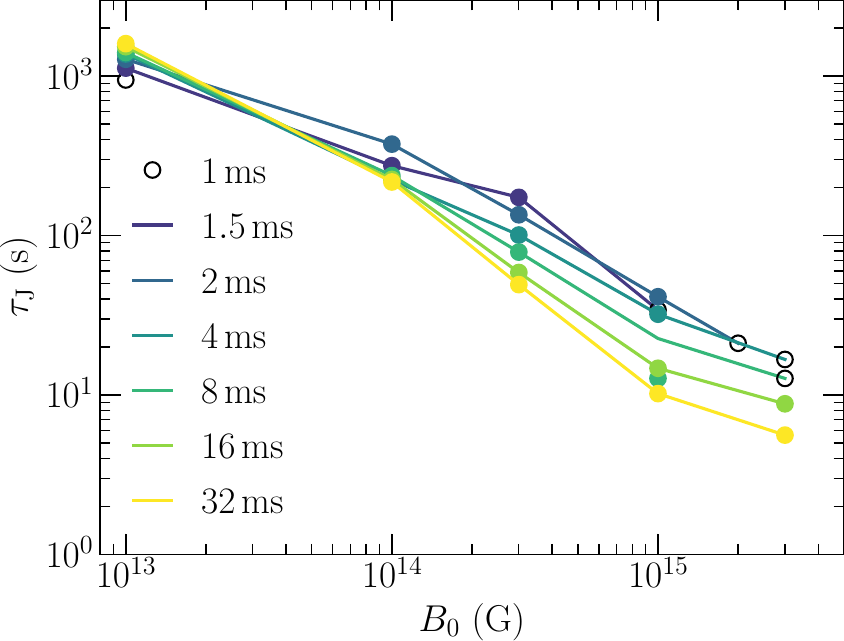}
\caption{Spindown time as a function of polar magnetic field strength and PNS spin period $P_{\star}$ in milliseconds at $L_{\rm \bar{\nu}_e}=8\times 10^{51}$\,ergs s$^{-1}$ for a 1.4\,M$_{\odot}$ PNS. Refer to the caption of Figure \ref{edot_B_P} for the meaning of filled and unfilled circles.}
\label{tauJ_B_P}
\end{figure}

\begin{figure*}
\centering
\includegraphics[width=\textwidth]{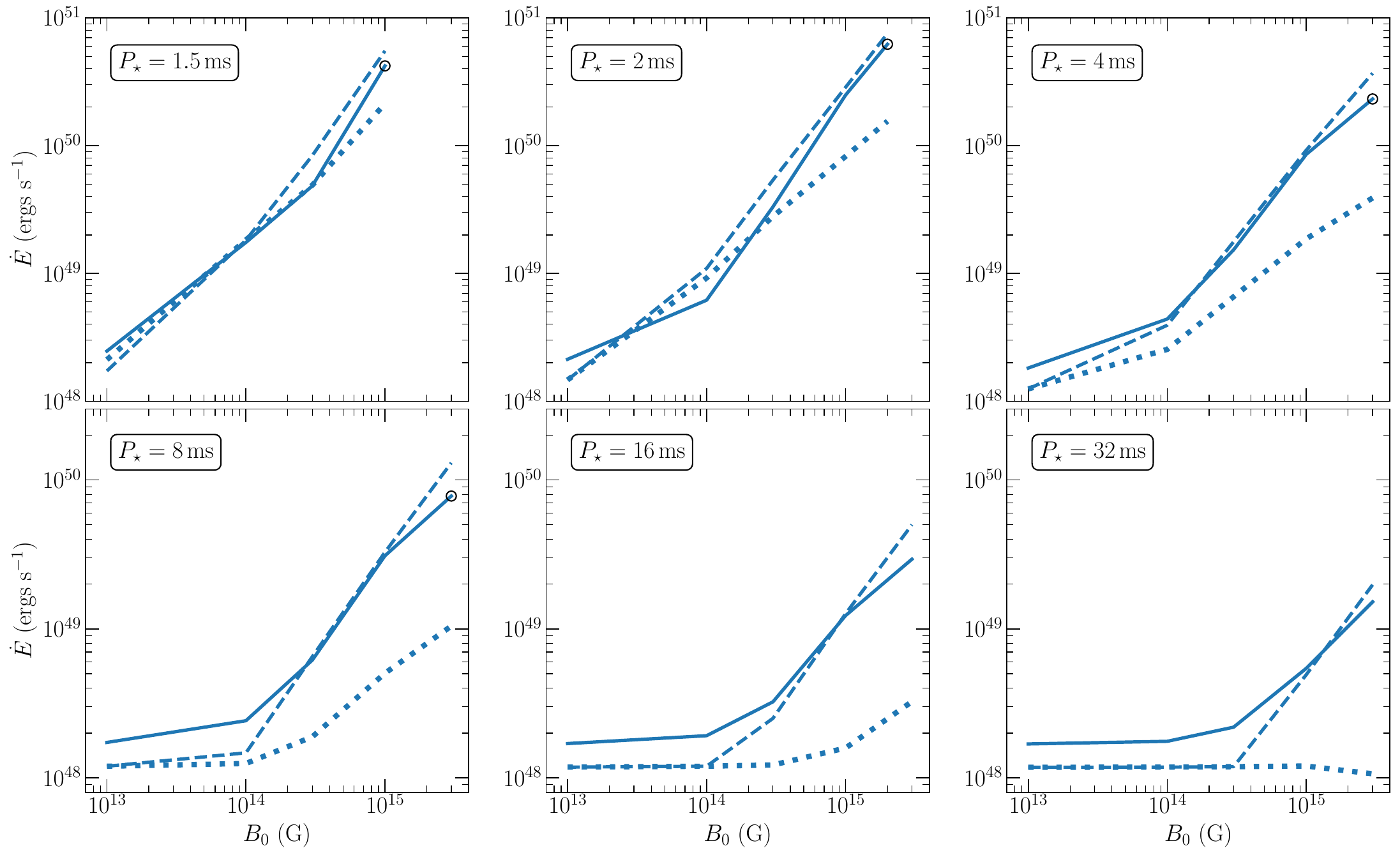}
\caption{Energy outflow rate $\dot{E}$ as a function of polar magnetic field strength $B_0$ and spin period $P_{\star}$ for a 1.4\,M$_{\odot}$ PNS at $L_{\rm \bar{\nu}_e}=8\times 10^{51}$\,ergs s$^{-1}$. The solid lines show $\dot{E}$ obtained from the simulations (see Table \ref{table1} and Figure \ref{edot_B_P}). The black unfilled circles on the solid lines represent results from potentially \textit{unreliable} simulations (see Section \ref{results} for details). The dashed lines are estimates of $\dot{E}$ assuming a monopole spindown ($\eta=2$) and the dotted lines are estimates assuming a dipole spindown ($\eta=3$) (see eqns. \ref{Ra_est} and \ref{Edot_est_expr}).}
\label{Edot_est}
\end{figure*}

\begin{figure*}
\centering{}
\includegraphics[width=\textwidth]{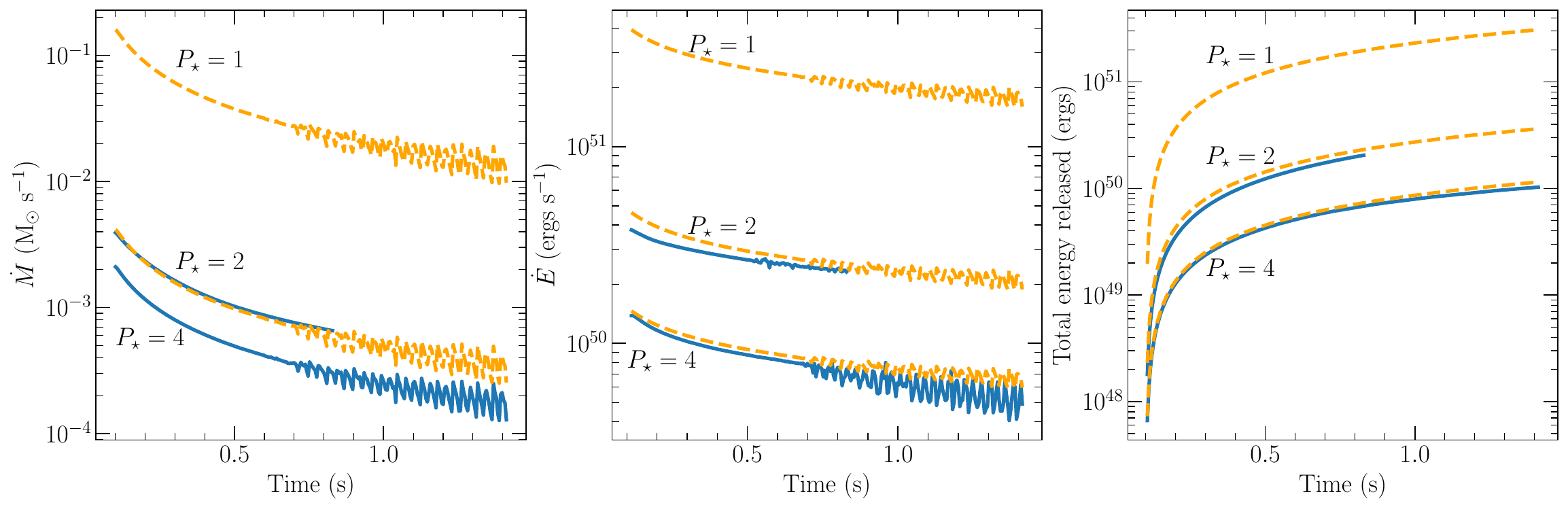}
\caption{Mass outflow rate, energy outflow rate and total energy released by the PNS as a function of time for a 1.4\,M$_{\odot}$ PNS at a polar magnetic field strength $B_0=10^{15}$\,G. The numbers next to the lines indicate the initial spin period of the PNS in milliseconds. The spin period of the PNS increases with time. The neutrino luminosity decreases with time according to the \citealt{Pons1999} cooling model. The start time of 0.1\,s on the time axis corresponds to the time at which $L_{\rm \bar{\nu}_e}=2\times10^{52}$\,ergs s$^{-1}$ in the \citealt{Pons1999} cooling model. The solid blue lines show data from the simulations. The dashed orange lines are $\dot{M}$ estimates (eqn.~\ref{mdot_est_expr}) and $\dot{E}$ estimates assuming monopole spindown ($\eta=2$, see eqs.~\ref{Ra_est} and \ref{Edot_est_expr})). We assume a constant PNS spin period for the estimates, which is a reasonable assumption (see Section \ref{results} for details). The total energy released is obtained by integrating $\dot{E}$ with time. We find that the estimates are very close to the results obtained from the simulations. $\dot{M}$ and $\dot{E}$ do not decrease monotonically after $\sim0.7$\,s due to the development of plasmoids.}
\label{time_evol_fig}
\end{figure*}

We do not attempt simulations with $P_{\star}<1.5$\,ms because of our density boundary condition at the inner edge (see eqn. \ref{densBC}). For $P_{\star}<1.5$\,ms, the base density at the equator is more than 30 times larger than the base density at the poles and hence, we cannot enforce our density boundary condition and net zero neutrino heating at the surface simultaneously. We just present one calculation at $P_{\star}=1$\,ms and $B_0=10^{13}$\,G as an estimate and emphasize that this result should not be trusted completely.

As mentioned in Section \ref{res}, more $\theta$ zones are required near the poles as $B_0$ and $\Omega_{\star}$ increase. We use $(N_r\,,\,N_{\theta})=(512\,,\,512)$ for simulations with $P_{\star}\leq4$\,ms and $B_0\geq10^{15}$\,G and $(N_r\,,\,N_{\theta})=(512\,,\,256)$ for the other calculations in Table \ref{table1}. Insufficient $\theta$ resolution near the poles results in the fluid pressure dropping below the degeneracy pressure (the least physically allowed pressure at a given density) near the outer boundary at $10000$\,km. This does not significantly affect the results as long as we use at least 256 $\theta$ zones. For example, at $P_{\star}=4$\,ms, $B_0=10^{15}$\,G and $L_{\rm \bar{\nu}_e}=8\times 10^{51}$\,ergs s$^{-1}$, time-averaged $\dot{J}$, $\dot{M}$ and $\dot{E}$ differ by less than 3\% between $(N_r\,,\,N_{\theta})=(512\,,\,256)$ and $(N_r\,,\,N_{\theta})=(512\,,\,512)$. All the physical quantities in Table \ref{table1} have been averaged over 0.5\,s run-time of the simulations. Ideally, we would like time-averages over $1-2$\,s, but these simulations are computationally expensive. To test if averaging over 0.5\,s is sufficient, we have run the simulation at $P_{\star}=4$\,ms, $B_0=10^{15}$\,G, $L_{\rm \bar{\nu}_e}=8\times 10^{51}$\,ergs s$^{-1}$ and $(N_r\,,\,N_{\theta})=(512\,,\,512)$ to 1\,s and we find that the 0.5\,s time average differs from the 1\,s time average  by less than 5\% .   

Magnetar strength magnetic fields aid in the efficient extraction of PNS rotational energy. Millisecond magnetars are reservoirs of large amount rotational kinetic energy:
\begin{align}
\label{Erot}
    E_{\rm rot}&=\frac{1}{2}I\Omega_{\star}^2 \\
    &\approx 3\times10^{52} {\rm ergs} \left(\frac{M_{\star}}{1.4 \,M_{\odot}}\right) \left(\frac{R_{\star}}{12\, \rm km}\right)^2 \left(\frac{P_{\star}}{1\, \rm ms}\right)^{-2},
\end{align}
where $M_{\star}$ is the mass and $I=\frac{2}{5}M_{\star}R_{\star}^2$ is the moment of inertia of the magnetar. The solid blue lines in Figure \ref{edot_B_P} show the asymptotic energy outflow rate as a function of polar magnetic field strength $B_0$ and rotation period $P_{\star}$. $\dot{E}$ increases with the polar magnetic field strength $B_0$ and rotation angular velocity $\Omega_{\star}$. As $B_0$ increases, the Alfv\'en radius increases and the wind effectively co-rotates with the PNS to larger radii, thus increasing $v_{\phi}$. This also results in a larger radial velocity $v_r$ due to centrifugal slinging, thus resulting in a larger $\dot{E}$. Faster rotation also increases $v_r$ and $v_{\phi}$ resulting in increasing $\dot{E}$ with $\Omega_{\star}$. The dashed orange lines show the $\dot{E} \propto B_0^{4/3}$ scaling starting from the highest value of $B_0$ for each spin period $P_{\star}$ (see the next subsection for scaling relations). We assume that $\dot{M}$ remains constant with $B_0$ in this scaling, which is not accurate (see Figure \ref{mdot_B_P}). However, we find that this scaling is close to the actual $\dot{E}$ profiles for $B_0\gtrsim 10^{14}$\,G and $P_{\star}\gtrsim4$\,ms. In the next subsection, we provide analytic estimates for $\dot{E}$ accounting for the variation of $\dot{M}$ with $B_0$ as well and show that those estimates approximate the actual $\dot{E}$ profiles better.

The maximum power emitted by a rotating vacuum dipole, which occurs for an orthogonal rotator, is given by $\dot{E}_{\rm d}=B_0^2\Omega_{\star}^4 R_{\star}^6/(6c^3)$. Figure \ref{edot_dip_rat} shows the ratio of energy outflow rate $\dot{E}$ obtained from our simulations to $\dot{E}_{\rm d}$.  In their first few seconds, proto-magnetars drive winds with energy loss rates that are many orders-of-magnitude larger than that given by the dipole formula. In general, $\dot{E}_{\rm d}$ should only be applied after the non-relativistic wind phase is over, many tens of seconds after PNS formation \citep{Thompson2004,Metzger2011}.

Figure \ref{mdot_B_P} shows the variation of the mass outflow rate $\dot{M}$ with $B_0$ and spin period $P_{\star}$. $\dot{M}$ increases as the PNS rotates faster at a fixed $B_0$ due to a larger centrifugal force. There are two competing effects that affect $\dot{M}$ as a function of $B_0$ at a fixed $P_{\star}$. First, as $B_0$ increases, the closed zone of the magnetic field gets bigger and more matter is trapped resulting in a smaller $\dot{M}$. Second, as the Alfv\'en radius increases due to increasing $B_0$ and the wind effectively co-rotates with the PNS to a larger radius, the centrifugal force results in a larger $\dot{M}$. These two competing effects cause the observed behavior of $\dot{M}$, that is, $\dot{M}$ increases with $B_0$ until the effect of centrifugal slinging dominates and then begins to decrease once the closed zone of the magnetic field dominates. Figure \ref{Rasph_B_P} shows the variation of the spherical Alfv\'en radius as a function of $B_0$ and $P_{\star}$. At a given value of $B_0$, the effective spherical Alfv\'en radius decreases as the PNS rotates faster due to larger wind speeds caused by magneto-centrifugal acceleration. As expected, at a fixed value of $P_{\star}$, $R_{\rm A}^{\rm sph}$ increases with $B_0$.

Figure \ref{tauJ_B_P} shows the spindown time $\tau_{\rm J}$ as a function of $B_0$ for different values of $P_{\star}$. At $B_0\lesssim 10^{13}$\,G, the magnetic field is too small to play a role in spindown and the spindown is mainly due to the centrifugal force. Hence, spindown time increases with spin period at low magnetic field strengths. $\tau_{\rm J}$ decreases with increasing $B_0$ at a fixed $P_{\star}$ due to larger angular momentum loss rate. At a given $B_0$, $\tau_{\rm J}$ in general decreases with increasing $P_{\star}$ as in \cite{Prasanna2022}. This is due to the fact that at high enough magnetic fields, spindown is dominated by the magnetic field as opposed to spin period. This means that $\dot{J}$ does not decrease proportionately with the total angular momentum $J$ of the PNS as the PNS spin period increases. However, this trend is broken at $B_0 \sim 10^{14}$\, G due to a transition between centrifugally dominated spindown and magnetically dominated spindown. We expect this trend to be broken again at high magnetic fields of $B_0 \gtrsim 10^{15}$\, G and very rapid rotation at $P_{\star}\lesssim 1.5$\,ms due to rapid increase of $\dot{J}$. We will explore this effect in a future work.

\subsection{Analytic estimates}
Since our inner density boundary condition (eqn. \ref{densBC}) does not allow us to simulate $P_{\star}<1.5$\,ms, we make use of analytics to estimate $\dot{E}$ and $\dot{M}$ for $P_{\star}<1.5$\,ms. We closely follow the estimates of \cite{Thompson2004}. We perform the analysis at the equator using the radial coordinate only. The radial component of the magnetic field can be approximated as,
\begin{equation}
    B_r(r)\approx\frac{B_0}{2}\left(\frac{R_{\star}}{r}\right)^{\eta},
\end{equation}
where $\eta=2$ for a ``split''-monopole field structure and $\eta=3$ for a dipole field structure. The factor of 2 in the denominator is to account for the fact that the magnitude of the magnetic field at the equator is half the magnitude at the pole for a dipole field. It is important to account for this because we label our simulations with the polar magnetic field strength $B_0$. At the radial Alfv\'en point $R_{\rm A}$, $B_r^2/8\pi=\rho v_r^2/2$. The density $\rho$ can be written in terms of the mass outflow rate. \cite{Thompson2004} assume spherical symmetry and use $\dot{M}=4\pi r^2\rho v_r$. But, $\dot{M}$ is not spherically symmetric in highly magnetized outflows. The ratio
\begin{equation}
Z=\frac{1}{\left(\rho v_r [\theta=\pi/2]\right)}\left(\int_0^{\pi}\rho v_r \sin \theta \ d\theta \right) 
\end{equation}
varies as a function of radius at which it is evaluated due to the structure of the magnetic field lines. The ratio also changes if the denominator is evaluated slightly ($\sim 1^{\circ}$) off the equator because of the equatorial current sheet in our axisymmetric calculations. The surface integral of $\dot{M}$ at a given radius can be written in terms of the quantities at the equator as $\dot{M}(r) \approx 2\pi Z r^2 \rho v_r$. This expression assumes a spherical surface while the Alfv\'en surface is actually cylindrical (see Figures \ref{vrvphi_B0} and \ref{vrvphi_Lnubar}). But this approximation is sufficient for our purposes. We find that $Z\approx 4$ fits our data well at the Alfv\'en radius.   
Hence, at $R_{\rm A}$,we have
\begin{equation}
    R_{\rm A}^{2\eta-2}\approx\frac{B_0^2}{2}R_{\star}^{2\eta}\dot{M}^{-1}v_{\rm A}^{-1} \ , 
\end{equation}
where $v_{\rm A}$ is the radial velocity at the Alfv\'en surface. From Figures \ref{vrvphi_B0}, \ref{vrvphi_Lnubar} and \ref{1d_th45}, we find that $v_r$ and $v_{\phi}$ are comparable when $R_{\rm A}\Omega_{\star}$ is much larger than the asymptotic radial velocity in the non-rotating non-magnetic (NRNM) limit (see also \citealt{Thompson2004}), which occurs as $\Omega_{\star}$ and $B_0$ increase. The asymptotic radial velocity $v_{\rm NRNM}$ varies with neutrino luminosity. We find from our 1D simulations that $v_{\rm NRNM}= 2\times 10^{9}$\,cm s$^{-1}$ at $L_{\rm \bar{\nu}_e}=8\times 10^{51}$\,ergs s$^{-1}$. Using $v_{\rm A}\approx R_{\rm A}\Omega_{\star}$, we get, 
\begin{equation}
    \label{Ra_est}
    R_{\rm A}^{2\eta-1}\approx\frac{B_0^2}{2}R_{\star}^{2\eta}\dot{M}^{-1} \Omega_{\star}^{-1}. 
\end{equation}
The asymptotic energy outflow rate can be estimated as,
\begin{equation}
\label{Edot_est_expr}
    \dot{E}_{\rm est}\approx \frac{1}{2}\dot{M}\left[{\rm \max} (R_{\rm A}\Omega_{\star}, v_{\rm NRNM}) \right]^2.
\end{equation}
where the values of $\dot{M}$ are obtained from the simulations. Figure \ref{Edot_est} shows $\dot{E}$ as a function of $B_0$ and $P_{\star}$. We plot the values obtained from the simulations (solid lines) and the monopole with $\eta=2$ (dashed lines) and dipole estimates with $\eta=3$ (dotted lines). We use the values of $\dot{M}$ from the simulations to obtain the $\dot{E}$ estimates.
We find that for rapid rotation, that is $P_{\star}\lesssim 8$\,ms, the monopole scalings match the simulation data very well. The monopole estimates are also good for 16\,ms and 32\,ms at $B_0 \gtrsim 10^{15}$\,G. As the PNS spin period increases and $B_0$ decreases, the estimates are less accurate because the assumption that $v_{\rm NRNM} \ll R_{\rm A} \Omega_{\star} $ breaks down. Although the magnetic field structure very close to the PNS surface is dipolar, the large scale magnetic field structure resembles a ``split''-monopole configuration as shown in Figures \ref{vrvphi_B0} and \ref{vrvphi_Lnubar}, because the pressure from the wind opens the magnetic field lines, resulting in good estimates assuming a monopole field configuration. 

We need to obtain estimates of $\dot{M}$ in order to estimate $\dot{E}$ for $P_{\star}<1.5$\,ms. From eqn. \ref{densBC}, we find that the density at the inner boundary depends exponentially on $\Omega_{\star}^2$. Hence, $\dot{M}\sim \exp\left(v_{\phi}^2/c_{\rm s}^2\right)$, where $c_{\rm s}$ is the adiabatic sound speed (see also \citealt{Thompson2004}). We have the following estimate relating $\dot{M}$ at different PNS rotation rates for a fixed value of $B_0$:
\begin{equation}
\label{mdot_est_expr}
    \dot{M} (\Omega_1)\approx \dot{M} (\Omega_2) \frac{\left(\int_0^{\pi}\exp\left[\frac{m_{\rm n}R_{\star}^2\Omega_{1}^2\sin^2\theta }{2k_{\rm B}T_0}\right] d\theta \right)} {\left(\int_0^{\pi}\exp\left[\frac{m_{\rm n}R_{\star}^2\Omega_{2}^2\sin^2\theta }{2k_{\rm B}T_0}\right] d\theta \right)},   
\end{equation}
where $T_0= 5$\,MeV is the approximate temperature at the PNS surface. At the PNS surface, $k_{\rm B}T_0/m_{\rm n} \approx c_{\rm s}^2$ for a base density $\sim 10^{12}$\,g cm$^{-3}$. As in eqn. \ref{densBC}, this expression assumes an EOS dominated by nucleons at the PNS surface, which is not appropriate at base densities $\lesssim 10^{10}$\,g cm$^{-3}$. 

\subsection{Evolutionary models}
In Figure \ref{time_evol_fig}, we present results from simulations with self-consistent evolution of $\Omega_{\star}$ and neutrino luminosity at $B_0=10^{15}$\,G for a 1.4\,M$_{\odot}$ PNS (see \citealt{Prasanna2022} for details about evolutionary models). We use the \cite{Pons1999} cooling model, starting from $L_{\rm \bar{\nu}_e}=2\times 10^{52}$\,ergs s$^{-1}$. We track the evolution down to $L_{\rm \bar{\nu}_e}=4\times 10^{51}$\,ergs s$^{-1}$ before the magnetosonic speeds exceed the speed of light. Although the $\dot{M}$ weighted average Alfv\'en speed is below $c$, we cannot continue the simulation because the equation of state cannot handle large magnetosonic speeds at a resolution of $(N_r\,,\,N_{\theta})=(512\,,\,512)$. Due to computational constraints, we do not attempt a simulation with a larger number of $\theta$ zones. Throughout the evolution, we hold the average neutrino energies constant at $\langle\epsilon_{\rm \bar{\nu}_e}\rangle=14$\,MeV and $\langle\epsilon_{\rm \nu_e}\rangle=11$\,MeV. In fact, \cite{Pons1999} find that average neutrino energies increase during the first $\sim5$\,s of evolution. Since the physics is uncertain at such early times after a supernova, we think it is better to get an underestimate. $\dot{E}$ increases with neutrino energy. The solid blue lines in Figure \ref{time_evol_fig} show the data from the simulations. The dashed orange lines are $\dot{M}$ estimates (eqn. \ref{mdot_est_expr}) and monopole ($\eta=2$) $\dot{E}$ estimates (eqns. \ref{Ra_est} and \ref{Edot_est_expr}). The total energy released by the PNS is computed as $\int_{t_0}^{t_1}\dot{E}dt$, where $dt$ is the time interval between consecutive output files of the simulation chosen to be 5\,ms and $t_0$ and $t_1$ are respectively the start and end time of neutrino luminosity evolution. At an initial spin period of $P_{\star}=4$\,ms and $B_0=10^{15}$\,G, we find that the PNS releases $10^{50}$\,ergs of energy during the first 1.5\,s of evolution. The PNS spins down to a period of 4.16\,ms during this time period of evolution. This spindown is a consequence of the PNS emitting $1.5\times10^{50}$\,ergs of rotational kinetic energy. This means that the total decrease in rotational kinetic energy of the PNS does not escape to large radii. A part of the rotational energy of the PNS is used to unbind the wind from the PNS. In contrast to slowly rotating magnetars (see \citealt{Prasanna2022}), the spin period of rapidly rotating magnetars does not evolve significantly during the cooling phase at the values of $B_0$ considered in this paper. Thus, we assume a constant spin period for the estimates shown by orange dashed lines in Figure \ref{time_evol_fig}, which is a reasonable assumption. As an example, a PNS with an initial spin period of 2\,ms spins down to a period of 2.05\,ms at the end of 0.8\,s of evolution at $B_0=10^{15}$\,G corresponding to a total energy release of $2\times 10^{50}$\,ergs.   

From Figure \ref{time_evol_fig}, we find that the estimates closely match the data from the simulations. Hence, we use these estimates to predict $\dot{M}$, $\dot{E}$ and total energy released at $P_{\star}=1$\,ms. The large increase in $\dot{M}$ at $P_{\star}=1$\,ms compared to slower rotation is evident. We find that a PNS starting with an initial spin period of 1\,ms releases about $3\times10^{51}$\,ergs and a PNS starting with an initial period of 2\,ms releases about $3.5\times10^{50}$\,ergs of energy during the first 1.5\,s of evolution at $B_0=10^{15}$\,G. As stated in the previous paragraph, we assume a constant spin period for the estimates. This slightly overestimates $\dot{E}$ and the total energy released by the PNS. Assuming a constant PNS spin period of 1.1\,ms (instead of 1\,ms) and 2.1\,ms (instead of 2\,ms) results in an energy yield of $2\times10^{51}$\,ergs and $3\times10^{50}$\,ergs respectively at the end of 1.5\,s at $B_0=10^{15}$\,G.    

\section{Discussion and Conclusions}
\label{conclusions}
We present a large set of 2D MHD simulations of proto-magnetars with a mass of 1.4\,$\rm M_{\odot}$, initial spin period $P_{\star}$ ranging from 1\,ms to 32\,ms and polar magnetic field strength $B_0$ ranging from $10^{13}$\,G to $3\times10^{15}$\,G at a fixed neutrino luminosity of $L_{\rm \bar{\nu}_e}=8\times 10^{51}$\,ergs s$^{-1}$ representative of the first few seconds after the explosion. We also present a couple of evolutionary models of a 1.4\,M$_{\odot}$ PNS to model the energy outflow from rapidly rotating millisecond magnetars. Figures \ref{edot_B_P} and \ref{mdot_B_P} show the variation of $\dot{E}$ and $\dot{M}$ respectively as a function of $B_0$ and spin period at a fixed neutrino luminosity. $\dot{E}$ increases rapidly with $B_0$, and with faster PNS rotation due to magneto-centrifugal slinging. $\dot{M}$ increases with faster rotation at a fixed $B_0$. As a function of $B_0$ at a fixed spin period, $\dot{M}$ first increases and then decreases. Effective co-rotation to a larger radius as $B_0$ increases and larger magneto-centrifugal acceleration of the subsonic wind atmosphere results in a larger $\dot{M}$, but the equatorial closed zone of the magnetic field near the PNS gets bigger as $B_0$ increases. A bigger closed zone traps more matter, leading to a decrease in $\dot{M}$. These two competing effects result in the observed behavior of $\dot{M}$. Figure \ref{tauJ_B_P} shows the variation of the spindown time $\tau_{\rm J}$ with $B_0$ and $P_{\star}$ at a fixed neutrino luminosity. $\tau_{\rm J}$ decreases with increasing $B_0$ at a fixed $P_{\star}$ due to larger $\dot{J}$. At a fixed $B_0$, $\tau_{\rm J}$ in general decreases with increasing spin period. Refer to the details in Section \ref{snapshots} for exceptions to this general trend. In contrast to `slowly' rotating magnetars with $P_{\star}\gtrsim 100$\,ms (see \citealt{Prasanna2022}), the spin period of rapidly rotating magnetars over the parameter range explored in this paper does not increase significantly during the cooling timescale.  

For the evolutionary models, we use $B_0=10^{15}$\,G and $P_{\star}\leq4$\,ms. We follow the evolution using the \cite{Pons1999} cooling model from $L_{\rm \bar{\nu}_e}=2\times 10^{52}$\,ergs s$^{-1}$ to $L_{\rm \bar{\nu}_e}=4\times 10^{51}$\,ergs s$^{-1}$. We assume a constant PNS radius during spindown. Figure \ref{time_evol_fig} shows the time evolution of $\dot{E}$, $\dot{M}$ and the total energy released by the magnetar. Since we are currently unable to simulate $P_{\star}<1.5$\,ms due to our density boundary condition at the PNS surface (see Section \ref{results} for details), we make use of some analytic estimates closely following \cite{Thompson2004}. Equations \ref{Ra_est} and \ref{Edot_est_expr} give estimates of the Alfv\'en radius and $\dot{E}$.  We test our analytic models against the simulations and find good agreement for the monopole ($\eta=2$) scaling. Figure \ref{Edot_est} compares $\dot{E}$ obtained from the simulations and the analytic estimates at a fixed neutrino luminosity. Figure \ref{time_evol_fig} compares the simulation data with the estimates of $\dot{E}$ and $\dot{M}$ for models with evolving neutrino luminosity and PNS spin period.

From the evolutionary models shown in Figure \ref{time_evol_fig}, we find that a PNS rotating at a period of 1\,ms releases $3\times10^{51}$\,ergs of energy during the first 1.5\,s of evolution at a polar magnetic field strength of $B_0=10^{15}$\,G. At $P_{\star}=2$\,ms and $B_0=10^{15}$\,G, a total of $3.5\times10^{50}$\,ergs of energy is released during the first 1.5\,s. Using eqn. \ref{Ra_est} and \ref{Edot_est_expr}, we find that the monopole scaling gives $\dot{E}\sim B_0^{4/3}$ (assuming $\dot{M}$ remains roughly the same). Thus, a PNS with an initial spin period of $P_{\star}=2$\,ms releases a total of $\sim1.5\times 10^{51}$\,ergs of energy at $B_0=3\times 10^{15}$\,G and $\sim 7.5\times 10^{51}$\,ergs at $B_0=10^{16}$\,G in the first 1.5\,s of evolution. A PNS with an initial spin period of $1$\,ms releases $\sim 1.3 \times 10^{52}$\,ergs and $6.5\times 10^{52}$\,ergs of energy at $B_0=3\times 10^{15}$\,G and $10^{16}$\,G respectively during the first 1.5\,s of evolution. The energy estimates at $B_0=10^{16}$\,G are comparable to or greater than the total rotational kinetic energy of the PNS because these estimates do not take into account the change in spin period of the PNS with time which is significant at fields $\sim 10^{16}$\,G. Although our modelling is different from \cite{Metzger2011} and we estimate the energy released by the PNS only during the first 1.5\,s, we observe good order of magnitude agreement with their results (see Figure 10 in \citealt{Metzger2011}). Based on the results and estimates of total energy released by rapidly rotating magnetars during the first $\sim 2$\,s of the cooling phase, it is plausible that sustained energy injection by magnetars through the relativistic wind phase can power GRBs. \cite{Metzger2011} argue that classical GRBs require magnetars with initial spin periods $\sim 1-2$\,ms and magnetic field strength $\sim 10^{16}$\,G. This is because high energy outflow rate $\dot{E}$ must be maintained during the highly relativistic wind phase as the PNS cools to setup production of GRBs.
\cite{Komissarov2007} and \cite{Bucciantini2008} demonstrate that the energy from neutrino driven winds of a magnetar can be channeled as a jet along the axis of rotation through strong toroidal magnetic fields far away from the magnetar. Although the magnetar injects energy into the surrounding environment at all values of the polar angle $\theta$, highly anisotropic jets can form along the poles  \citep{Komissarov2007,Bucciantini2008}. We will explore the effects of interaction between the magnetar outflow and the surrounding medium in a future work.

Observations of X-ray plateaus associated with GRBs offer evidence for a millisecond magnetar central engine (see \citealt{Lyons2010} for example). \cite{Lyons2010} find that the limits on the magnetic field strength and spin period of the PNS obtained from the luminosity and duration of the plateaus are consistent with those of millisecond magnetars. However, \cite{Beniamini2021} argue that association of plateaus with magnetars formed in a binary neutron star merger requires a lot of fine-tuning and magnetars may not be a natural explanation for the plateaus. 

The energy in the anisotropic jet before break out from the progenitor star may significantly affect the energy associated with the supernova itself. Based on the velocity of the material in the jet, a varying fraction of the total energy released by the magnetar goes into powering a GRB (see \citealt{Komissarov2007} for example). \cite{Margalit2018} argue that based on the misalignment angle between the magnetic and rotation axes of the PNS, a varying fraction of the magnetar's spin down power goes into powering an energetic SN and the remaining power is used to power a jet. In such a scenario, it is possible that the magnetar transfers $\geq 10^{51}$\,ergs of energy to the SN itself, resulting in a hyper-energetic supernova (``hypernova''). Assuming that the angle between the rotation and magnetic axes does not significantly affect $\dot{E}$, our results suggest that extreme magnetars with spin periods $1-2$\,ms and $B_0\gtrsim 10^{15}-10^{16}$\,G may power hypernovae. 

Millisecond magnetars with more moderate ($\sim 10^{14}$\,G) magnetic field strength do not have enough power to drive a GRB, but these can power SLSNe. Magnetars with polar magnetic field strength $B_0\lesssim 5\times10^{14}$\,G do not expend most of their rotational energy during the non-relativistic wind phase probed by the calculations presented in this paper. From Figure \ref{time_evol_fig}, we find that a PNS rotating at a period of 2\,ms releases about $3.5\times10^{50}$\,ergs of energy at a polar magnetic field strength of $B_0=10^{15}$\,G during the first 1.5\,s of evolution. From eqns. \ref{Ra_est} and \ref{Edot_est_expr}, we find that monopole scalings yield $\dot{E}\sim B_0^{4/3}$. Thus, at $P_{\star}=2$\,ms and $B_0=10^{14}$\,G, the PNS releases just about $1.8\times10^{49}$\,ergs of energy during the first 1.5\,s of evolution. At 2\,ms, a magnetar is a reservoir of $7.5\times10^{51}$\,ergs of rotational kinetic energy (see eqn. \ref{Erot}). Figures \ref{edot_B_P} and \ref{tauJ_B_P} and the above estimate show that magnetars with moderate magnetic fields do not spindown significantly and do not release most of their rotational energy during the early wind phase. Hence, they have sufficient rotational energy that can be released during the magnetic dipole radiation phase. Such magnetars, after the non-relativistic and subsequently relativistic wind phase, can release their rotational energy through magnetic dipole radiation on timescales of days to weeks and contribute significantly towards brightening the supernova light curve. \cite{Kasen2010} find that magnetars with $P_{\star}\sim2-20$\,ms and moderate $B_0$ can explain the light curves of SLSNe. \cite{Metzger2015} find that such magnetars with long spindown time can explain both luminous supernovae and ultra long GRBs with a duration of $\gtrsim 10^{4}$\,s. \cite{Metzger2018_2} speculate that late time accretion might allow both SLSNe and GRBs to be powered by magnetars with the same surface fields.

As shown in Figures \ref{vrvphi_B0} and \ref{vrvphi_Lnubar} and \cite{Prasanna2022} and as predicted in \cite{Thompson2003}, quasi-periodic plasmoid eruptions occur near the equator for certain combinations of polar magnetic field strength $B_0$ and neutrino luminosity when the magnetic energy density exceeds the thermal pressure of the wind at the equator (see \citealt{Thompson2018}). Plasmoids contain high entropy material and thermodynamically favorable conditions for $r$-process nucleosynthesis if the wind is neutron-rich, or the $rp$- or $\nu p$-process if the medium is proton-rich \citep{Pruet2006,Thompson2018}. Comparing our results here with the plasmoids found in \cite{Prasanna2022}, suggests that the rotation period of the PNS affects the maximum entropy achievable in the plasmoids and their dynamics. We will focus on aspects of nucleosynthesis in a future work. The asymptotic value of the electron fraction $Y_{\rm e}$ decreases with faster PNS rotation due to magneto-centrifugal slinging as the material gets advected away faster \citep{Metzger2008_2}. This effect is more pronounced with spin periods $\lesssim 1$\,ms and magnetic fields $\gtrsim 10^{15}$\,G (see Figure 2 in \citealt{Metzger2008_2}). We observe about 1\% reduction in asymptotic $Y_{\rm e}$ at $P_{\star}=2$\,ms compared to $P_{\star}=32$\,ms at $B_0=10^{15}$\,G. We will focus on $P_{\star}\lesssim 1$\,ms in a future work. 

As in \cite{Prasanna2022}, we use non-relativistic physics in our simulations. This limits the maximum value of the polar magnetic field strength $B_0$ in our simulations. As $B_0$ increases and/or as the neutrino luminosity decreases, the magnetosonic speeds approach the speed of light. Without including relativistic effects, we cannot follow the evolution of the PNS throughout the cooling phase that lasts $\sim10-100$\,s. We plan to include relativistic effects and follow the PNS evolution throughout the cooling phase in a future work.

\section*{Acknowledgments}
\label{section:acknowledgements}
We thank the referee for useful comments and suggestions that helped us to improve this paper. We thank Brian Metzger for useful comments on this paper. We thank Yan-Fei Jiang, Zhaohuan Zhu, Jim Stone, Kengo Tomida, and Adam Finley for helpful discussions. TAT thanks Asif ud-Doula, Brian Metzger, Phil Chang, Niccol\'o Bucciantini, and Eliot Quataert for discussions and collaboration on this and related topics. TP, TAT, and MJR are supported in part by NASA grant 80NSSC20K0531. MC acknowledges support from the U.~S.\ National Science Foundation (NSF) under Grants AST-1714267 and PHY-1804048 (the latter via the Max-Planck/Princeton Center (MPPC) for Plasma Physics). We have run our simulations on the Ohio supercomputer \citep{OhioSupercomputerCenter1987}. Parts of the results in this work make use of the colormaps in the CMasher package \citep{CMasher}.

\section*{Data Availability}
The implementation of the EOS and the problem generator file to run the simulations using Athena++ are available upon request.


\bibliographystyle{mnras}
\bibliography{ref} 

\begin{thebibliography}{}
\makeatletter
\relax
\def\mn@urlcharsother{\let\do\@makeother \do\$\do\&\do\#\do\^\do\_\do\%\do\~}
\def\mn@doi{\begingroup\mn@urlcharsother \@ifnextchar [ {\mn@doi@}
  {\mn@doi@[]}}
\def\mn@doi@[#1]#2{\def\@tempa{#1}\ifx\@tempa\@empty \href
  {http://dx.doi.org/#2} {doi:#2}\else \href {http://dx.doi.org/#2} {#1}\fi
  \endgroup}
\def\mn@eprint#1#2{\mn@eprint@#1:#2::\@nil}
\def\mn@eprint@arXiv#1{\href {http://arxiv.org/abs/#1} {{\tt arXiv:#1}}}
\def\mn@eprint@dblp#1{\href {http://dblp.uni-trier.de/rec/bibtex/#1.xml}
  {dblp:#1}}
\def\mn@eprint@#1:#2:#3:#4\@nil{\def\@tempa {#1}\def\@tempb {#2}\def\@tempc
  {#3}\ifx \@tempc \@empty \let \@tempc \@tempb \let \@tempb \@tempa \fi \ifx
  \@tempb \@empty \def\@tempb {arXiv}\fi \@ifundefined
  {mn@eprint@\@tempb}{\@tempb:\@tempc}{\expandafter \expandafter \csname
  mn@eprint@\@tempb\endcsname \expandafter{\@tempc}}}

\bibitem[\protect\citeauthoryear{{Barkov} \& {Komissarov}}{{Barkov} \&
  {Komissarov}}{2008}]{Barkov2008}
{Barkov} M.~V.,  {Komissarov} S.~S.,  2008, \mn@doi [\mnras]
  {10.1111/j.1745-3933.2008.00427.x}, \href
  {https://ui.adsabs.harvard.edu/abs/2008MNRAS.385L..28B} {385, L28}

\bibitem[\protect\citeauthoryear{{Beniamini} \& {Lu}}{{Beniamini} \&
  {Lu}}{2021}]{Beniamini2021}
{Beniamini} P.,  {Lu} W.,  2021, \mn@doi [\apj] {10.3847/1538-4357/ac1678},
  \href {https://ui.adsabs.harvard.edu/abs/2021ApJ...920..109B} {920, 109}

\bibitem[\protect\citeauthoryear{{Bucciantini}, {Quataert}, {Arons}, {Metzger}
  \& {Thompson}}{{Bucciantini} et~al.}{2007}]{Bucciantini2007}
{Bucciantini} N.,  {Quataert} E.,  {Arons} J.,  {Metzger} B.~D.,   {Thompson}
  T.~A.,  2007, \mn@doi [\mnras] {10.1111/j.1365-2966.2007.12164.x}, \href
  {https://ui.adsabs.harvard.edu/abs/2007MNRAS.380.1541B} {380, 1541}

\bibitem[\protect\citeauthoryear{{Bucciantini}, {Quataert}, {Arons}, {Metzger}
  \& {Thompson}}{{Bucciantini} et~al.}{2008}]{Bucciantini2008}
{Bucciantini} N.,  {Quataert} E.,  {Arons} J.,  {Metzger} B.~D.,   {Thompson}
  T.~A.,  2008, \mn@doi [\mnras] {10.1111/j.1745-3933.2007.00403.x}, \href
  {https://ui.adsabs.harvard.edu/abs/2008MNRAS.383L..25B} {383, L25}

\bibitem[\protect\citeauthoryear{{Bucciantini}, {Metzger}, {Thompson}  \&
  {Quataert}}{{Bucciantini} et~al.}{2012}]{Bucciantini2012}
{Bucciantini} N.,  {Metzger} B.~D.,  {Thompson} T.~A.,   {Quataert} E.,  2012,
  \mn@doi [\mnras] {10.1111/j.1365-2966.2011.19810.x}, \href
  {https://ui.adsabs.harvard.edu/abs/2012MNRAS.419.1537B} {419, 1537}

\bibitem[\protect\citeauthoryear{{Burrows} \& {Mazurek}}{{Burrows} \&
  {Mazurek}}{1982}]{Burrows1982}
{Burrows} A.,  {Mazurek} T.~J.,  1982, \mn@doi [\apj] {10.1086/160169}, \href
  {https://ui.adsabs.harvard.edu/abs/1982ApJ...259..330B} {259, 330}

\bibitem[\protect\citeauthoryear{Center}{Center}{1987}]{OhioSupercomputerCenter1987}
Center O.~S.,  1987, Ohio Supercomputer Center, \url
  {http://osc.edu/ark:/19495/f5s1ph73}

\bibitem[\protect\citeauthoryear{{Coleman}}{{Coleman}}{2020}]{Coleman2020}
{Coleman} M. S.~B.,  2020, \mn@doi [\apjs] {10.3847/1538-4365/ab82ff}, \href
  {https://ui.adsabs.harvard.edu/abs/2020ApJS..248....7C} {248, 7}

\bibitem[\protect\citeauthoryear{{De Luca}, {Caraveo}, {Mereghetti}, {Tiengo}
  \& {Bignami}}{{De Luca} et~al.}{2006}]{Luca2006}
{De Luca} A.,  {Caraveo} P.~A.,  {Mereghetti} S.,  {Tiengo} A.,   {Bignami}
  G.~F.,  2006, \mn@doi [Science] {10.1126/science.1129185}, \href
  {https://ui.adsabs.harvard.edu/abs/2006Sci...313..814D} {313, 814}

\bibitem[\protect\citeauthoryear{{Desai}, {Siegel}  \& {Metzger}}{{Desai}
  et~al.}{2022}]{Desai2022}
{Desai} D.,  {Siegel} D.~M.,   {Metzger} B.~D.,  2022, \mn@doi [\apj]
  {10.3847/1538-4357/ac69da}, \href
  {https://ui.adsabs.harvard.edu/abs/2022ApJ...931..104D} {931, 104}

\bibitem[\protect\citeauthoryear{{Dessart}, {Burrows}, {Livne}  \&
  {Ott}}{{Dessart} et~al.}{2008}]{Dessart2008}
{Dessart} L.,  {Burrows} A.,  {Livne} E.,   {Ott} C.~D.,  2008, \mn@doi [\apjl]
  {10.1086/527519}, \href
  {https://ui.adsabs.harvard.edu/abs/2008ApJ...673L..43D} {673, L43}

\bibitem[\protect\citeauthoryear{{Duncan} \& {Thompson}}{{Duncan} \&
  {Thompson}}{1992}]{Duncan1992}
{Duncan} R.~C.,  {Thompson} C.,  1992, \mn@doi [\apjl] {10.1086/186413}, \href
  {https://ui.adsabs.harvard.edu/abs/1992ApJ...392L...9D} {392, L9}

\bibitem[\protect\citeauthoryear{{Duncan}, {Shapiro}  \& {Wasserman}}{{Duncan}
  et~al.}{1986}]{Duncan1986}
{Duncan} R.~C.,  {Shapiro} S.~L.,   {Wasserman} I.,  1986, \mn@doi [\apj]
  {10.1086/164587}, \href
  {https://ui.adsabs.harvard.edu/abs/1986ApJ...309..141D} {309, 141}

\bibitem[\protect\citeauthoryear{{Gottlieb}, {Liska}, {Tchekhovskoy},
  {Bromberg}, {Lalakos}, {Giannios}  \& {M{\"o}sta}}{{Gottlieb}
  et~al.}{2022}]{Gottlieb2022}
{Gottlieb} O.,  {Liska} M.,  {Tchekhovskoy} A.,  {Bromberg} O.,  {Lalakos} A.,
  {Giannios} D.,   {M{\"o}sta} P.,  2022, \mn@doi [\apjl]
  {10.3847/2041-8213/ac7530}, \href
  {https://ui.adsabs.harvard.edu/abs/2022ApJ...933L...9G} {933, L9}

\bibitem[\protect\citeauthoryear{{Granot}, {Guetta}  \& {Gill}}{{Granot}
  et~al.}{2017}]{Granot2017}
{Granot} J.,  {Guetta} D.,   {Gill} R.,  2017, \mn@doi [\apjl]
  {10.3847/2041-8213/aa991d}, \href
  {https://ui.adsabs.harvard.edu/abs/2017ApJ...850L..24G} {850, L24}

\bibitem[\protect\citeauthoryear{{Janka} \& {Mueller}}{{Janka} \&
  {Mueller}}{1996}]{Janka1996}
{Janka} H.~T.,  {Mueller} E.,  1996, \aap, \href
  {https://ui.adsabs.harvard.edu/abs/1996A&A...306..167J} {306, 167}

\bibitem[\protect\citeauthoryear{{Kasen} \& {Bildsten}}{{Kasen} \&
  {Bildsten}}{2010}]{Kasen2010}
{Kasen} D.,  {Bildsten} L.,  2010, \mn@doi [\apj]
  {10.1088/0004-637X/717/1/245}, \href
  {https://ui.adsabs.harvard.edu/abs/2010ApJ...717..245K} {717, 245}

\bibitem[\protect\citeauthoryear{{King}, {Pringle}  \& {Wickramasinghe}}{{King}
  et~al.}{2001}]{King2001}
{King} A.~R.,  {Pringle} J.~E.,   {Wickramasinghe} D.~T.,  2001, \mn@doi
  [\mnras] {10.1046/j.1365-8711.2001.04184.x}, \href
  {https://ui.adsabs.harvard.edu/abs/2001MNRAS.320L..45K} {320, L45}

\bibitem[\protect\citeauthoryear{{Klebesadel}, {Strong}  \&
  {Olson}}{{Klebesadel} et~al.}{1973}]{Klebesadel1973}
{Klebesadel} R.~W.,  {Strong} I.~B.,   {Olson} R.~A.,  1973, \mn@doi [\apjl]
  {10.1086/181225}, \href
  {https://ui.adsabs.harvard.edu/abs/1973ApJ...182L..85K} {182, L85}

\bibitem[\protect\citeauthoryear{{Komissarov} \& {Barkov}}{{Komissarov} \&
  {Barkov}}{2007}]{Komissarov2007}
{Komissarov} S.~S.,  {Barkov} M.~V.,  2007, \mn@doi [\mnras]
  {10.1111/j.1365-2966.2007.12485.x}, \href
  {https://ui.adsabs.harvard.edu/abs/2007MNRAS.382.1029K} {382, 1029}

\bibitem[\protect\citeauthoryear{{Lai} \& {Qian}}{{Lai} \&
  {Qian}}{1998}]{Lai1998}
{Lai} D.,  {Qian} Y.-Z.,  1998, \mn@doi [\apj] {10.1086/306203}, \href
  {https://ui.adsabs.harvard.edu/abs/1998ApJ...505..844L} {505, 844}

\bibitem[\protect\citeauthoryear{{Levan}, {Wynn}, {Chapman}, {Davies}, {King},
  {Priddey}  \& {Tanvir}}{{Levan} et~al.}{2006}]{Levan2006}
{Levan} A.~J.,  {Wynn} G.~A.,  {Chapman} R.,  {Davies} M.~B.,  {King} A.~R.,
  {Priddey} R.~S.,   {Tanvir} N.~R.,  2006, \mn@doi [\mnras]
  {10.1111/j.1745-3933.2006.00144.x}, \href
  {https://ui.adsabs.harvard.edu/abs/2006MNRAS.368L...1L} {368, L1}

\bibitem[\protect\citeauthoryear{Lyons, O'Brien, Zhang, Willingale, Troja  \&
  Starling}{Lyons et~al.}{2010}]{Lyons2010}
Lyons N.,  O'Brien P.~T.,  Zhang B.,  Willingale R.,  Troja E.,   Starling R.
  L.~C.,  2010, \mn@doi [Monthly Notices of the Royal Astronomical Society]
  {10.1111/j.1365-2966.2009.15538.x}, 402, 705

\bibitem[\protect\citeauthoryear{{MacFadyen} \& {Woosley}}{{MacFadyen} \&
  {Woosley}}{1999}]{MacFadyen1999}
{MacFadyen} A.~I.,  {Woosley} S.~E.,  1999, \mn@doi [\apj] {10.1086/307790},
  \href {https://ui.adsabs.harvard.edu/abs/1999ApJ...524..262M} {524, 262}

\bibitem[\protect\citeauthoryear{{Margalit}, {Metzger}, {Thompson}, {Nicholl}
  \& {Sukhbold}}{{Margalit} et~al.}{2018}]{Margalit2018}
{Margalit} B.,  {Metzger} B.~D.,  {Thompson} T.~A.,  {Nicholl} M.,   {Sukhbold}
  T.,  2018, \mn@doi [\mnras] {10.1093/mnras/sty013}, \href
  {https://ui.adsabs.harvard.edu/abs/2018MNRAS.475.2659M} {475, 2659}

\bibitem[\protect\citeauthoryear{{Metzger}, {Thompson}  \&
  {Quataert}}{{Metzger} et~al.}{2007}]{Metzger2007}
{Metzger} B.~D.,  {Thompson} T.~A.,   {Quataert} E.,  2007, \mn@doi [\apj]
  {10.1086/512059}, \href
  {https://ui.adsabs.harvard.edu/abs/2007ApJ...659..561M} {659, 561}

\bibitem[\protect\citeauthoryear{{Metzger}, {Quataert}  \&
  {Thompson}}{{Metzger} et~al.}{2008a}]{Metzger2008}
{Metzger} B.~D.,  {Quataert} E.,   {Thompson} T.~A.,  2008a, \mn@doi [\mnras]
  {10.1111/j.1365-2966.2008.12923.x}, \href
  {https://ui.adsabs.harvard.edu/abs/2008MNRAS.385.1455M} {385, 1455}

\bibitem[\protect\citeauthoryear{{Metzger}, {Thompson}  \&
  {Quataert}}{{Metzger} et~al.}{2008b}]{Metzger2008_2}
{Metzger} B.~D.,  {Thompson} T.~A.,   {Quataert} E.,  2008b, \mn@doi [\apj]
  {10.1086/526418}, \href
  {https://ui.adsabs.harvard.edu/abs/2008ApJ...676.1130M} {676, 1130}

\bibitem[\protect\citeauthoryear{{Metzger}, {Giannios}, {Thompson},
  {Bucciantini}  \& {Quataert}}{{Metzger} et~al.}{2011}]{Metzger2011}
{Metzger} B.~D.,  {Giannios} D.,  {Thompson} T.~A.,  {Bucciantini} N.,
  {Quataert} E.,  2011, \mn@doi [\mnras] {10.1111/j.1365-2966.2011.18280.x},
  \href {https://ui.adsabs.harvard.edu/abs/2011MNRAS.413.2031M} {413, 2031}

\bibitem[\protect\citeauthoryear{{Metzger}, {Margalit}, {Kasen}  \&
  {Quataert}}{{Metzger} et~al.}{2015}]{Metzger2015}
{Metzger} B.~D.,  {Margalit} B.,  {Kasen} D.,   {Quataert} E.,  2015, \mn@doi
  [\mnras] {10.1093/mnras/stv2224}, \href
  {https://ui.adsabs.harvard.edu/abs/2015MNRAS.454.3311M} {454, 3311}

\bibitem[\protect\citeauthoryear{{Metzger}, {Thompson}  \&
  {Quataert}}{{Metzger} et~al.}{2018a}]{Metzger2018}
{Metzger} B.~D.,  {Thompson} T.~A.,   {Quataert} E.,  2018a, \mn@doi [\apj]
  {10.3847/1538-4357/aab095}, \href
  {https://ui.adsabs.harvard.edu/abs/2018ApJ...856..101M} {856, 101}

\bibitem[\protect\citeauthoryear{{Metzger}, {Beniamini}  \&
  {Giannios}}{{Metzger} et~al.}{2018b}]{Metzger2018_2}
{Metzger} B.~D.,  {Beniamini} P.,   {Giannios} D.,  2018b, \mn@doi [\apj]
  {10.3847/1538-4357/aab70c}, \href
  {https://ui.adsabs.harvard.edu/abs/2018ApJ...857...95M} {857, 95}

\bibitem[\protect\citeauthoryear{{M{\"o}sta} et~al.,}{{M{\"o}sta}
  et~al.}{2014}]{Mosta2014}
{M{\"o}sta} P.,  et~al., 2014, \mn@doi [\apjl] {10.1088/2041-8205/785/2/L29},
  \href {https://ui.adsabs.harvard.edu/abs/2014ApJ...785L..29M} {785, L29}

\bibitem[\protect\citeauthoryear{{M{\"o}sta}, {Radice}, {Haas}, {Schnetter}  \&
  {Bernuzzi}}{{M{\"o}sta} et~al.}{2020}]{Mosta2020}
{M{\"o}sta} P.,  {Radice} D.,  {Haas} R.,  {Schnetter} E.,   {Bernuzzi} S.,
  2020, \mn@doi [\apjl] {10.3847/2041-8213/abb6ef}, \href
  {https://ui.adsabs.harvard.edu/abs/2020ApJ...901L..37M} {901, L37}

\bibitem[\protect\citeauthoryear{{Nicholl}, {Guillochon}  \&
  {Berger}}{{Nicholl} et~al.}{2017}]{Nicholl2017}
{Nicholl} M.,  {Guillochon} J.,   {Berger} E.,  2017, \mn@doi [\apj]
  {10.3847/1538-4357/aa9334}, \href
  {https://ui.adsabs.harvard.edu/abs/2017ApJ...850...55N} {850, 55}

\bibitem[\protect\citeauthoryear{{Nomoto} \& {Kondo}}{{Nomoto} \&
  {Kondo}}{1991}]{Nomoto1991}
{Nomoto} K.,  {Kondo} Y.,  1991, \mn@doi [\apjl] {10.1086/185922}, \href
  {https://ui.adsabs.harvard.edu/abs/1991ApJ...367L..19N} {367, L19}

\bibitem[\protect\citeauthoryear{{Olausen} \& {Kaspi}}{{Olausen} \&
  {Kaspi}}{2014}]{Olausen2014}
{Olausen} S.~A.,  {Kaspi} V.~M.,  2014, \mn@doi [\apjs]
  {10.1088/0067-0049/212/1/6}, \href
  {https://ui.adsabs.harvard.edu/abs/2014ApJS..212....6O} {212, 6}

\bibitem[\protect\citeauthoryear{{Otsuki}, {Tagoshi}, {Kajino}  \&
  {Wanajo}}{{Otsuki} et~al.}{2000}]{Otsuki2000}
{Otsuki} K.,  {Tagoshi} H.,  {Kajino} T.,   {Wanajo} S.-y.,  2000, \mn@doi
  [\apj] {10.1086/308632}, \href
  {https://ui.adsabs.harvard.edu/abs/2000ApJ...533..424O} {533, 424}

\bibitem[\protect\citeauthoryear{{Pons}, {Reddy}, {Prakash}, {Lattimer}  \&
  {Miralles}}{{Pons} et~al.}{1999}]{Pons1999}
{Pons} J.~A.,  {Reddy} S.,  {Prakash} M.,  {Lattimer} J.~M.,   {Miralles}
  J.~A.,  1999, \mn@doi [\apj] {10.1086/306889}, \href
  {https://ui.adsabs.harvard.edu/abs/1999ApJ...513..780P} {513, 780}

\bibitem[\protect\citeauthoryear{{Prasanna}, {Coleman}, {Raives}  \&
  {Thompson}}{{Prasanna} et~al.}{2022}]{Prasanna2022}
{Prasanna} T.,  {Coleman} M. S.~B.,  {Raives} M.~J.,   {Thompson} T.~A.,  2022,
  \mn@doi [\mnras] {10.1093/mnras/stac2651}, \href
  {https://ui.adsabs.harvard.edu/abs/2022MNRAS.517.3008P} {517, 3008}

\bibitem[\protect\citeauthoryear{{Pruet}, {Hoffman}, {Woosley}, {Janka}  \&
  {Buras}}{{Pruet} et~al.}{2006}]{Pruet2006}
{Pruet} J.,  {Hoffman} R.~D.,  {Woosley} S.~E.,  {Janka} H.~T.,   {Buras} R.,
  2006, \mn@doi [\apj] {10.1086/503891}, \href
  {https://ui.adsabs.harvard.edu/abs/2006ApJ...644.1028P} {644, 1028}

\bibitem[\protect\citeauthoryear{{Qian} \& {Woosley}}{{Qian} \&
  {Woosley}}{1996}]{QW1996}
{Qian} Y.~Z.,  {Woosley} S.~E.,  1996, \mn@doi [\apj] {10.1086/177973}, \href
  {https://ui.adsabs.harvard.edu/abs/1996ApJ...471..331Q} {471, 331}

\bibitem[\protect\citeauthoryear{{Raives}, {Coleman}  \& {Thompson}}{{Raives}
  et~al.}{2023}]{Raives2023}
{Raives} M.~J.,  {Coleman} M. S.~B.,   {Thompson} T.~A.,  2023, \mn@doi [arXiv
  e-prints] {10.48550/arXiv.2302.05462}, \href
  {https://ui.adsabs.harvard.edu/abs/2023arXiv230205462R} {p. arXiv:2302.05462}

\bibitem[\protect\citeauthoryear{{Ricci} et~al.,}{{Ricci}
  et~al.}{2021}]{Ricci2021}
{Ricci} R.,  et~al., 2021, \mn@doi [\mnras] {10.1093/mnras/staa3241}, \href
  {https://ui.adsabs.harvard.edu/abs/2021MNRAS.500.1708R} {500, 1708}

\bibitem[\protect\citeauthoryear{{Rowlinson}, {Gompertz}, {Dainotti},
  {O'Brien}, {Wijers}  \& {van der Horst}}{{Rowlinson}
  et~al.}{2014}]{Rowlinson2014}
{Rowlinson} A.,  {Gompertz} B.~P.,  {Dainotti} M.,  {O'Brien} P.~T.,  {Wijers}
  R.~A.~M.~J.,   {van der Horst} A.~J.,  2014, \mn@doi [\mnras]
  {10.1093/mnras/stu1277}, \href
  {https://ui.adsabs.harvard.edu/abs/2014MNRAS.443.1779R} {443, 1779}

\bibitem[\protect\citeauthoryear{{Scheck}, {Kifonidis}, {Janka}  \&
  {M{\"u}ller}}{{Scheck} et~al.}{2006}]{Scheck2006}
{Scheck} L.,  {Kifonidis} K.,  {Janka} H.~T.,   {M{\"u}ller} E.,  2006, \mn@doi
  [\aap] {10.1051/0004-6361:20064855}, \href
  {https://ui.adsabs.harvard.edu/abs/2006A&A...457..963S} {457, 963}

\bibitem[\protect\citeauthoryear{{Stanek} et~al.,}{{Stanek}
  et~al.}{2003}]{Stanek2003}
{Stanek} K.~Z.,  et~al., 2003, \mn@doi [\apjl] {10.1086/376976}, \href
  {https://ui.adsabs.harvard.edu/abs/2003ApJ...591L..17S} {591, L17}

\bibitem[\protect\citeauthoryear{{Stone}, {Tomida}, {White}  \&
  {Felker}}{{Stone} et~al.}{2019}]{Athena++}
{Stone} J.~M.,  {Tomida} K.,  {White} C.,   {Felker} K.~G.,  2019, {Athena++:
  Radiation GR magnetohydrodynamics code} (\mn@eprint {ascl} {1912.005})

\bibitem[\protect\citeauthoryear{{Thompson}}{{Thompson}}{2003}]{Thompson2003}
{Thompson} T.~A.,  2003, \mn@doi [\apjl] {10.1086/374261}, \href
  {https://ui.adsabs.harvard.edu/abs/2003ApJ...585L..33T} {585, L33}

\bibitem[\protect\citeauthoryear{{Thompson} \& {Duncan}}{{Thompson} \&
  {Duncan}}{1993}]{Thompson1993}
{Thompson} C.,  {Duncan} R.~C.,  1993, \mn@doi [\apj] {10.1086/172580}, \href
  {https://ui.adsabs.harvard.edu/abs/1993ApJ...408..194T} {408, 194}

\bibitem[\protect\citeauthoryear{{Thompson} \& {ud-Doula}}{{Thompson} \&
  {ud-Doula}}{2018}]{Thompson2018}
{Thompson} T.~A.,  {ud-Doula} A.,  2018, \mn@doi [\mnras]
  {10.1093/mnras/sty480}, \href
  {https://ui.adsabs.harvard.edu/abs/2018MNRAS.476.5502T} {476, 5502}

\bibitem[\protect\citeauthoryear{{Thompson}, {Burrows}  \& {Meyer}}{{Thompson}
  et~al.}{2001}]{Thompson2001}
{Thompson} T.~A.,  {Burrows} A.,   {Meyer} B.~S.,  2001, \mn@doi [\apj]
  {10.1086/323861}, \href
  {https://ui.adsabs.harvard.edu/abs/2001ApJ...562..887T} {562, 887}

\bibitem[\protect\citeauthoryear{{Thompson}, {Chang}  \& {Quataert}}{{Thompson}
  et~al.}{2004}]{Thompson2004}
{Thompson} T.~A.,  {Chang} P.,   {Quataert} E.,  2004, \mn@doi [\apj]
  {10.1086/421969}, \href
  {https://ui.adsabs.harvard.edu/abs/2004ApJ...611..380T} {611, 380}

\bibitem[\protect\citeauthoryear{{Timmes} \& {Swesty}}{{Timmes} \&
  {Swesty}}{2000}]{Timmes2000}
{Timmes} F.~X.,  {Swesty} F.~D.,  2000, \mn@doi [\apjs] {10.1086/313304}, \href
  {https://ui.adsabs.harvard.edu/abs/2000ApJS..126..501T} {126, 501}

\bibitem[\protect\citeauthoryear{{Usov}}{{Usov}}{1992}]{Usov1992}
{Usov} V.~V.,  1992, \mn@doi [\nat] {10.1038/357472a0}, \href
  {https://ui.adsabs.harvard.edu/abs/1992Natur.357..472U} {357, 472}

\bibitem[\protect\citeauthoryear{{Vidotto}, {Jardine}, {Morin}, {Donati},
  {Opher}  \& {Gombosi}}{{Vidotto} et~al.}{2014}]{Vidotto2014}
{Vidotto} A.~A.,  {Jardine} M.,  {Morin} J.,  {Donati} J.~F.,  {Opher} M.,
  {Gombosi} T.~I.,  2014, \mn@doi [\mnras] {10.1093/mnras/stt2265}, \href
  {https://ui.adsabs.harvard.edu/abs/2014MNRAS.438.1162V} {438, 1162}

\bibitem[\protect\citeauthoryear{{Vlasov}, {Metzger}  \& {Thompson}}{{Vlasov}
  et~al.}{2014}]{Vlasov2014}
{Vlasov} A.~D.,  {Metzger} B.~D.,   {Thompson} T.~A.,  2014, \mn@doi [\mnras]
  {10.1093/mnras/stu1667}, \href
  {https://ui.adsabs.harvard.edu/abs/2014MNRAS.444.3537V} {444, 3537}

\bibitem[\protect\citeauthoryear{{Vlasov}, {Metzger}, {Lippuner}, {Roberts}  \&
  {Thompson}}{{Vlasov} et~al.}{2017}]{Vlasov2017}
{Vlasov} A.~D.,  {Metzger} B.~D.,  {Lippuner} J.,  {Roberts} L.~F.,
  {Thompson} T.~A.,  2017, \mn@doi [\mnras] {10.1093/mnras/stx478}, \href
  {https://ui.adsabs.harvard.edu/abs/2017MNRAS.468.1522V} {468, 1522}

\bibitem[\protect\citeauthoryear{{Wanajo}}{{Wanajo}}{2006}]{Wanajo2006}
{Wanajo} S.,  2006, \mn@doi [\apj] {10.1086/505483}, \href
  {https://ui.adsabs.harvard.edu/abs/2006ApJ...647.1323W} {647, 1323}

\bibitem[\protect\citeauthoryear{{Wanajo}, {Kajino}, {Mathews}  \&
  {Otsuki}}{{Wanajo} et~al.}{2001}]{Wanajo2001}
{Wanajo} S.,  {Kajino} T.,  {Mathews} G.~J.,   {Otsuki} K.,  2001, \mn@doi
  [\apj] {10.1086/321339}, \href
  {https://ui.adsabs.harvard.edu/abs/2001ApJ...554..578W} {554, 578}

\bibitem[\protect\citeauthoryear{{White}, {Burrows}, {Coleman}  \&
  {Vartanyan}}{{White} et~al.}{2022}]{White2022}
{White} C.~J.,  {Burrows} A.,  {Coleman} M. S.~B.,   {Vartanyan} D.,  2022,
  \mn@doi [\apj] {10.3847/1538-4357/ac4507}, \href
  {https://ui.adsabs.harvard.edu/abs/2022ApJ...926..111W} {926, 111}

\bibitem[\protect\citeauthoryear{{Woosley}}{{Woosley}}{1993}]{Woosley1993}
{Woosley} S.~E.,  1993, \mn@doi [\apj] {10.1086/172359}, \href
  {https://ui.adsabs.harvard.edu/abs/1993ApJ...405..273W} {405, 273}

\bibitem[\protect\citeauthoryear{{Woosley}}{{Woosley}}{2010}]{Woosley2010}
{Woosley} S.~E.,  2010, \mn@doi [\apjl] {10.1088/2041-8205/719/2/L204}, \href
  {https://ui.adsabs.harvard.edu/abs/2010ApJ...719L.204W} {719, L204}

\bibitem[\protect\citeauthoryear{{Woosley} \& {Bloom}}{{Woosley} \&
  {Bloom}}{2006}]{Woosley2006}
{Woosley} S.~E.,  {Bloom} J.~S.,  2006, \mn@doi [\araa]
  {10.1146/annurev.astro.43.072103.150558}, \href
  {https://ui.adsabs.harvard.edu/abs/2006ARA&A..44..507W} {44, 507}

\bibitem[\protect\citeauthoryear{{Zhang} \& {M{\'e}sz{\'a}ros}}{{Zhang} \&
  {M{\'e}sz{\'a}ros}}{2001}]{Zhang2001}
{Zhang} B.,  {M{\'e}sz{\'a}ros} P.,  2001, \mn@doi [\apjl] {10.1086/320255},
  \href {https://ui.adsabs.harvard.edu/abs/2001ApJ...552L..35Z} {552, L35}

\bibitem[\protect\citeauthoryear{{van Putten} \& {Della Valle}}{{van Putten} \&
  {Della Valle}}{2023}]{Van_putten2023}
{van Putten} M. H.~P.~M.,  {Della Valle} M.,  2023, \mn@doi [\aap]
  {10.1051/0004-6361/202142974}, \href
  {https://ui.adsabs.harvard.edu/abs/2023A&A...669A..36V} {669, A36}

\bibitem[\protect\citeauthoryear{van~der Velden}{van~der
  Velden}{2020}]{CMasher}
van~der Velden E.,  2020, \mn@doi [Journal of Open Source Software]
  {10.21105/joss.02004}, 5, 2004

\makeatother
\end{thebibliography}

\bsp	
\label{lastpage}
\end{document}